\documentclass[%
 reprint,
unsortedaddress,
 amsmath,amssymb,
 aps,
]{revtex4-2}

\usepackage[english]{babel}  
\usepackage{epsfig} 
\usepackage{graphicx}
\usepackage{dcolumn}
\usepackage{bm}
\usepackage[colorlinks=true]{hyperref}

\begin{document}

\preprint{APS/123-QED}

\title{Evolution of Phonon Transport Across Structural Phase Transitions in MgAgSb}

\author{Luman Shang}
\affiliation{Advanced Thermal Management Technology and Functional Materials Laboratory, Ministry of Education Key Laboratory of NSLSCS, School of Energy and Mechanical Engineering, Nanjing Normal University, Nanjing 210023, P. R. China}

\author{Yu Wu}
\email{wuyu@njnu.edu.cn}
\affiliation{Advanced Thermal Management Technology and Functional Materials Laboratory, Ministry of Education Key Laboratory of NSLSCS, School of Energy and Mechanical Engineering, Nanjing Normal University, Nanjing 210023, P. R. China}

\author{Yufan Liu}
\affiliation{Advanced Thermal Management Technology and Functional Materials Laboratory, Ministry of Education Key Laboratory of NSLSCS, School of Energy and Mechanical Engineering, Nanjing Normal University, Nanjing 210023, P. R. China}

\author{Shuming Zeng}
\affiliation{College of Physics Science and Technology, Yangzhou University, Jiangsu, Yangzhou 225009, China}

\author{Gang Tang}
\affiliation{School of Interdisciplinary Science, Beijing Institute of Technology, Beijing 100081, China}

\author{Chenhan Liu}
\email{chenhanliu@njnu.edu.cn}
\affiliation{Advanced Thermal Management Technology and Functional Materials Laboratory, Ministry of Education Key Laboratory of NSLSCS, School of Energy and Mechanical Engineering, Nanjing Normal University, Nanjing 210023, P. R. China}
%
%
%
%
%
%

\begin{abstract}
MgAgSb, a promising thermoelectric material, undergoes reversible phase transitions that drastically alter its thermal transport behavior. Using first-principles calculations, we systematically investigate the lattice thermal conductivity ($\kappa_L$) of its three phases: $\alpha$, $\beta$, and $\gamma$, revealing a progressive increase following $\alpha < \beta < \gamma$. This trend originates from distinct scattering mechanisms. Four-phonon scattering substantially suppresses the particle-like conductivity ($\kappa_p$) in the $\beta$ and $\gamma$ phases, while electron-phonon scattering provides a minor additional reduction. In contrast, the wave-like conductivity ($\kappa_c$) from coherent phonon tunneling is highest in the complex $\alpha$ phase, contributing up to 44\% of $\kappa_L$. Notably, the temperature dependence of $\kappa_L$ differs fundamentally between phases: in $\beta$, the weak $\kappa_p$ variation arises from a decreasing Grüneisen parameter with temperature; in $\alpha$, the strong rise in $\kappa_c$ with temperature counteracts the decay of $\kappa_p$. Our findings establish a comprehensive picture of thermal transport in MgAgSb, highlighting the phase-dependent interplay between particle-like and wave-like phonon contributions.

\end{abstract}

\maketitle
\newcommand{\Rmnum}[1]{\uppercase\expandafter{\romannumeral#1}}
\section*{\Rmnum{1}.INTRODUCTION}
Understanding lattice thermal transport across structural phase transitions is important for both fundamental phonon physics and the optimization of thermoelectric (TE) materials\cite{Wang2023c,Jiang2018a,Yu2025,Shi2020,Chebli2025}. In crystalline solids, heat conduction is commonly described within the Peierls--Boltzmann framework, where phonons are treated as particle-like quasiparticles characterized by well-defined group velocities and lifetimes\cite{Omini1995,Yang2018,Li2021}. This description works well for simple crystals with weakly overlapping phonon branches, but it can become insufficient in structurally complex materials, strongly anharmonic systems, and compounds with dense low-lying optical modes. In such cases, heat transport may also involve couplings between different vibrational eigenstates, giving rise to a coherent contribution beyond the conventional phonon-gas picture\cite{Luckyanova2012,Wang2005,Colvard1980,Zeng2025}. The unified Wigner formulation provides a natural framework for this regime by decomposing the lattice thermal conductivity ($\kappa_L$) into a particle-like term ($\kappa_p$) and a wave-like coherent term ($\kappa_c$)\cite{Simoncelli2022,Legenstein2025,Isaeva2019}, thereby connecting the Peierls limit of simple crystals with the Allen--Feldman-like regime of complex and glasslike solids. In recent years, this framework has proved especially useful for materials with ultralow or weakly temperature-dependent thermal conductivity ($\kappa$)\cite{Mukhopadhyay2018,Shenogin2009,Zhang2021,Das2023}.

At present, accurate first-principles predictions for thermal transport increasingly require going beyond the traditional three phonon (3ph) approximation. When lower-order channels are limited or anharmonicity is strong, four phonon (4ph) scattering has been proven to produce significant corrections\cite{Yang2019,Xia2020,Feng2018}. A reliable description of the phase-dependent $\kappa$ requires simultaneously addressing higher-order anharmonic scattering\cite{Ravichandran2020,Ji2024,Tadano2015a} and coherent off-diagonal transport\cite{Simoncelli2019,Isaeva2019}. This has been confirmed in the study of the multiphase CsPbBr$_3$, where considering 4ph instead of only 3ph results in a 40\% reduction in $\kappa_L$\cite{Wang2023}. Meanwhile, in metallic or semimetallic systems, electron-phonon (el-ph) scattering can become another decay pathway for the phonons that thermal transport\cite{Liao2015,Jain2016,Wang2016,Yang2002}. For example, in TaN, after considering el-ph scattering, the $\kappa_L$ decreases by 15.6\%\cite{Kundu2021,Wijeyesekera1984}.

MgAgSb is an excellent material to address this question\cite{Zhao2014a,Kirkham2012,Li2025,Xie2023a}. As a near-room-temperature TE material, $\alpha$-MgAgSb exhibits favorable electrical transport together with intrinsically low $\kappa_L$, which has been associated with its distorted crystal structure, weak hierarchical bonding, and suppression of transverse acoustic phonons\cite{Liu2018a,Huang2023,Back2025,Ying2015}. MgAgSb also undergoes successive structural transitions from the low-temperature $\alpha$ phase to the intermediate $\beta$ phase and then to the high-temperature $\gamma$ phase, accompanied by substantial changes in symmetry, disorder, and electronic structure. In particular, the $\alpha$ phase is semiconducting\cite{Li2025}. And theoretical calculations of the $\beta$ and $\gamma$ phases in their stable structures show that they are metallic\cite{Mi2017}, implying that el-ph scattering may become relevant in these phases. Although previous studies have clarified the crystal structures and TE properties of individual MgAgSb phases, and have identified microscopic origins of the ultralow $\kappa$ in $\alpha$-MgAgSb, a unified picture of heat transport across the $\alpha$--$\beta$--$\gamma$ sequence is still lacking, especially with respect to the interplay among $\kappa_p$, $\kappa_c$, 4ph scattering, and el-ph scattering.

In this work, we systematically investigate the $\kappa_L$ of $\alpha$-, $\beta$-, and $\gamma$-MgAgSb from first principles within a unified thermal-transport framework. By combining the $\kappa_p$ and the $\kappa_c$, and by explicitly incorporating 3ph, 4ph, and el-ph scattering where appropriate, we establish a comprehensive microscopic picture of phase-dependent heat transport in this system. We show that 4ph scattering strongly suppresses $\kappa_p$ in the $\beta$ and $\gamma$ phases, while el-ph scattering provides an additional reduction. By contrast, the coherent contribution is most significant in the structurally complex $\alpha$ phase, where it accounts for a remarkably large fraction of $\kappa_L$. We further show that the temperature dependence of $\kappa_L$ differs qualitatively among the three phases: the weak variation of $\kappa_p$ in the $\beta$ phase is linked to a reduction in the Gr\"uneisen parameter with increasing temperature, whereas in the $\alpha$ phase the growth of $\kappa_c$ partly compensates the decay of $\kappa_p$, leading to an unusually weak overall temperature dependence. These results clarify the microscopic origin of thermal transport across phase transitions in MgAgSb and highlight the need to treat particle-like transport, coherent vibrational coupling, higher-order anharmonicity, and el-ph scattering on an equal footing in phase-change TE materials.


\section*{\Rmnum{2}.COMPUTATIONAL METHODS}

First-principles calculations were performed using the Vienna Ab Initio Simulation Package (VASP) based on density functional theory (DFT)\cite{Kresse1996a}, employing the projector augmented-wave (PAW) method with the PBEsol exchange-correlation functional\cite{Perdew2008}. A plane-wave cutoff energy of 520 eV was used. Atomic positions are optimized with an energy convergence criterion of $10^{-8}$ eV between consecutive steps and a maximum Hellmann-Feynman force tolerance of $10^{-6}$ eV/\r{A}. The primitive cells of the $\alpha$ (24 atoms), $\beta$ (6 atoms), and $\gamma$ (3 atoms) phases were optimized using $\Gamma$-centered $\mathbf{q}$-meshes of $5\times5\times5$, $8\times8\times5$, and $9\times9\times9$, respectively.

For the $ab$ $initio$ molecular dynamics (AIMD) simulations, supercells containing 192 atoms were constructed for each phase by expanding their respective optimized primitive cells: $2\times2\times2$ supercell for the $\alpha$ phase, $4\times4\times2$ supercell for the $\beta$ phase, and $4\times4\times4$ supercell for the $\gamma$ phase. The AIMD simulations were run for 20 ps with a 1 fs timestep, employing a $\Gamma$-centered $1\times1\times1$ $\mathbf{k}$-mesh. The temperature-dependent effective potential (TDEP)\cite{Hellman2013,Knoop2024} method was then used to extract the second-, third-, and fourth-order interatomic force constants (IFCs). The cutoff radii for the second-, third-, and fourth-order IFCs were set to 10, 6, and 3~\AA~for the $\alpha$ phase, 10, 6, and 4~\AA~for the $\beta$ phase, and 10, 6, and 5~\AA~for the $\gamma$ phase.

The lattice thermal conductivity ($\kappa_L$) was computed by solving the Boltzmann transport equation as implemented in the ShengBTE software\cite{Li2014,Han2022}. A $\mathbf{q}$-mesh of $9\times9\times9$, $15\times15\times15$, and $21\times21\times21$ was adopted in the first irreducible Brillouin zone for the $\alpha$, $\beta$, and $\gamma$ phases, respectively, and with a Gaussian smearing width of 0.1. For the four-phonon (4ph) scattering rates were efficiently calculated by using the maximum likelihood estimation method with a sample size of $3\times10^5$\cite{Guo2024}. Furthermore, the influence of electron-phonon (el-ph) scattering for the $\beta$ and $\gamma$ phases was evaluated using the EPW code\cite{Ponce2016}, alone with QUANTUM ESPRESSO package\cite{Giannozzi2009}. In this work, the fully relativistic norm-conserving pseudopotentials that include the spin-orbit coupling (SOC) are employed\cite{Hamann2013}. A plane-wave energy cutoff of 100 Ry was adopted. For the Brillouin zone of MgAgSb, initial coarse grids of $8\times8\times8$ for $k$- and $4\times4\times4$ for $q$-points were used. These were subsequently interpolated onto dense $50\times50\times50$ grids via maximally localized Wannier functions. The \( \kappa_{\text{ph}} \) was computed using the ShengBTE package\cite{Ji2024,Li2014,Li2012b,Li2012c}, which was modified to incorporate both el-ph and 4ph scattering processes.

\section*{\Rmnum{3}.RESULTS AND DISCUSSION}

\begin{figure*}[ht!]
\centering
\epsfig{file=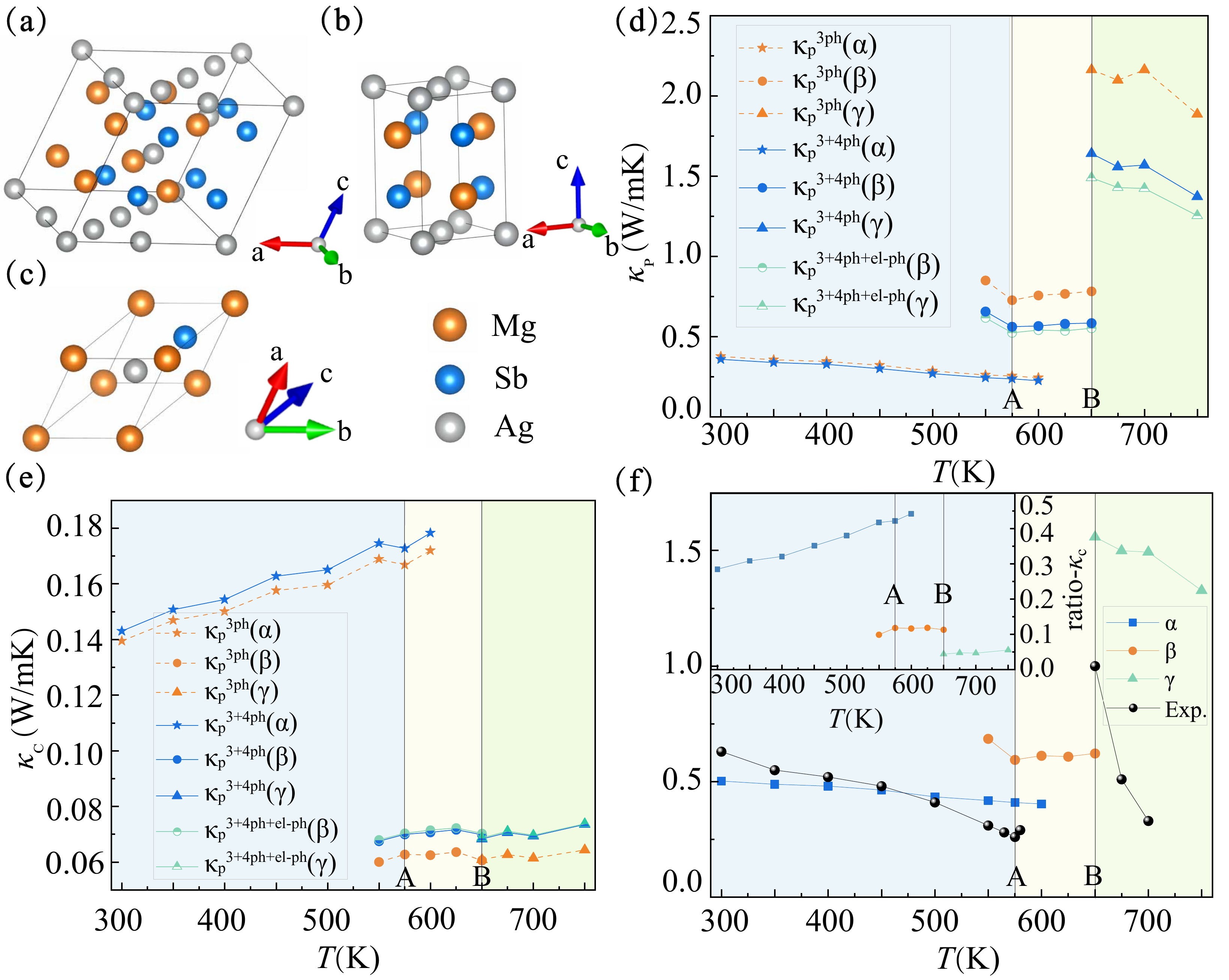, width=0.8\textwidth}
\caption{(a-c) The primitive cell of the $\alpha$ ,the $\beta$, the $\gamma$ phases of MgAgSb. The orange, blue, and silver colors represent Mg, Sb, and Ag atoms. The calculated (d) particle-like thermal conductivity ($\kappa_p$) and (e) the wave-like thermal conductivity ($\kappa_c$) for the three phases under 3ph, 3+4ph, 3+4ph+el-ph scattering. (f) Compares the calculated lattice thermal conductivity ($\kappa_L$) with experimental data, and the proportion of $\kappa_c$ to $\kappa_L$. Phase transition point: A (575 K),B (650 K).}
\label{Fig1}
\end{figure*}

Figures 1(a)-(c) illustrate the three primitive cell structures of MgAgSb within its operating temperature range. The low-temperature $\alpha$ phase (below $\sim 560 K$) crystallizes in a tetragonal structure with space group $I\text{-}4c2$, containing 24 atoms—the most complex structure among the three phases. The intermediate $\beta$ phase ($\sim 560–650 K$) also adopts a tetragonal structure (space group $P4/nmm$) with 6 atoms. The high-temperature $\gamma$ phase (above $\sim 650 K$) possesses a cubic structure (space group $F\text{-}43m$), consisting of only 3 atoms and having the highest symmetry\cite{Huang2023}.

Figure 1(d) presents the calculated particle-like thermal conductivity ($\kappa_p$) for the three phases under different scattering mechanisms. For the $\alpha$ phase, the inclusion of four-phonon (4ph) scattering has a negligible effect on $\kappa_p$, and its temperature dependence follows $\kappa_p \sim T^{-0.66}$. For the $\beta$ phase, 4ph scattering reduces $\kappa_p$ by 22.8\% at 575 K compared to three-phonon (3ph) scattering alone, and additional electron-phonon (el-ph) scattering further reduces $\kappa_p$ by 6.7\%. Meanwhile, the $\beta$ phase exhibits weak temperature dependence with a slight increase in $\kappa_p$ between 575 and 650 K. For the $\gamma$ phase, 4ph scattering reduces $\kappa_p$ by 24.2\% at 650 K compared to 3ph scattering alone, and el-ph scattering leads to an additional 9.2\% reduction. Overall, $\kappa_p$ follows the order $\alpha < \beta < \gamma$.

Figure 1(e) shows the wave-like thermal conductivity ($\kappa_c$) as a function of temperature under different scattering mechanisms. $\kappa_c$ generally increases with temperature for all phases. The $\alpha$ phase exhibits the most pronounced variation, rising from 0.14 W/mK at 300 K to 0.18 W/mK at 550 K—an increase of 28.6\% at 3+4ph. In contrast, $\kappa_c$ of the $\beta$ and $\gamma$ phases shows much weaker temperature dependence. For the $\beta$ phase at 575 K, 4ph scattering increases $\kappa_c$ by 11.5\% compared to 3ph scattering alone, while el-ph scattering has a negligible effect. Similarly, for the $\gamma$ phase at 650 K, 4ph scattering increases $\kappa_c$ by 12.4\% compared to 3ph scattering alone, with el-ph scattering again showing little impact.

\begin{figure*}[ht!]
\centering
\epsfig{file=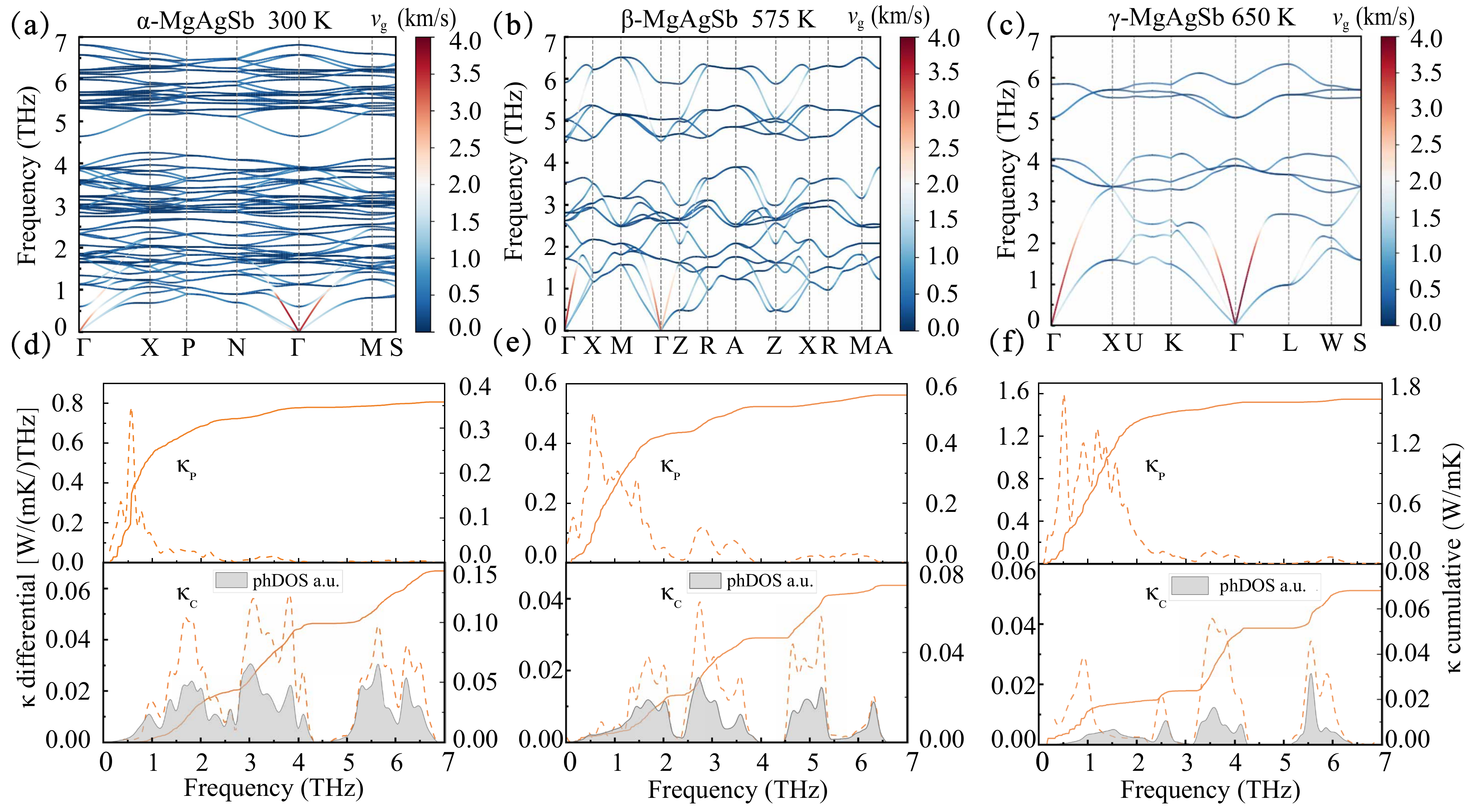, width=1\textwidth}
\caption{The phonon group velocity ($v_g$) projected onto the phonon dispersions of (a) $\alpha$ , (b) $\beta$, and (c) $\gamma$ at 300 K, 575 K, 650 K, respectively. The cumulative and differential $\kappa_p$ and $\kappa_c$ as a function of phonon frequency and the phonon density of states (phDOS) of (d) $\alpha$, (e) $\beta$, and (f) $\gamma$ at 300 K, 575 K, 650 K, respectively.}
\label{Fig2}
\end{figure*}

Figure 1(f) compares the calculated lattice thermal conductivity ($\kappa_L$) with experimental data. For the $\alpha$ phase, the calculated $\kappa_L$ (3,4ph) agrees well with experimental values\cite{Huang2023}. Due to the weak temperature dependence of $\kappa_p$ and the strong temperature dependence of $\kappa_c$, the overall $\kappa_L$ exhibits a reduced temperature exponent, with $\kappa_L \sim T^{-0.33}$\cite{Ouyang2025}. For the $\beta$ phase, $\kappa_L^{\text{3+4ph+el-ph}}$ shows a slight increase between 575 and 650 K, which arises from the intrinsic temperature dependence of the $\beta$ phase. The experimental $\kappa_L$ exhibits an upturn near 550–560 K, corresponding to the emergence of the $\beta$ phase. For the $\gamma$ phase, the calculated $\kappa_L^{\text{3+4ph+el-ph}}$ at 650 K is 1.56 W/mK, notably higher than the experimental value of ~1 W/mK. The deviation may mainly stems from the presence of impurities in the experimental samples and the slow $\beta \to \gamma$ phase transition process which is in a unbalanced state. Moreover, the theoretical calculations are based on an ideal pure phase and do not fully consider the influence of the phase transition\cite{Huang2023}. The inset in Figure 1(f) shows the proportion of $\kappa_c$ to $\kappa_L$. Notably, in the $\alpha$ phase, $\kappa_c$ contributes up to 44.3\% of the total $\kappa_L$, consistent with the view that wave-like phonon transport plays a significant role in low-$\kappa$ materials\cite{Zhao2025,Zheng2024}.

To further analyze the values and differences of $\kappa_L$ across the three phases, figure 2 presents an analysis of the phonon transmission characteristic, with all data calculated at each phase's respective temperature ($\alpha$ at 300 K, $\beta$ at 575 K, and $\gamma$ at 650 K). Figures 2(a)-(c) show the phonon group velocity ($v_g$) projected onto the phonon dispersions. The $\alpha$ phase exhibits the densest phonon spectrum and the softest acoustic branches, with a maximum acoustic frequency of 1.2 THz. Its acoustic $v_g$ remains around 0.5 km/s over a broad frequency range (1-7 THz). The $\beta$ phase shows a sparser spectrum, with acoustic branches extending to 1.9 THz and higher $v_g$ values of about 1.0 km/s within the range of 2 to 7 THz. The $\gamma$ phase, with only 3 atoms per cell, has the simplest spectrum, featuring acoustic branches up to 3.3 THz and $v_g$ of 1.5 km/s above 3 THz. Figures 2(d)–(f) display the cumulative and differential $\kappa_p$ and $\kappa_c$ as a function of phonon frequency for the three phases at their respective temperatures. For all phases, $\kappa_p$ is predominantly contributed by low-frequency acoustic modes, where $v_g$ is higher and dispersion is stronger. In contrast, $\kappa_c$ is governed by phonon modes that contribute minimally to $\kappa_p$, primarily those located in high-phDOS regions. The differential of $\kappa_c$ closely resembles the phonon density of states (phDOS) curve for all three phases. The flat and dense phonon spectrum of the $\alpha$ phase enables a large number of phonon pairs with small frequency differences ($\omega-\omega'$), thereby giving rise to its substantial $\kappa_c$. For the $\beta$ and $\gamma$ phases, the phonon spectra become sparse, the number of coherent phonons with a smaller frequency difference decreases\cite{Wu2023b}. Consequently, their $\kappa_c$ values are much smaller than that of the $\alpha$ phase.

\begin{figure*}[ht!]
\centering
\epsfig{file=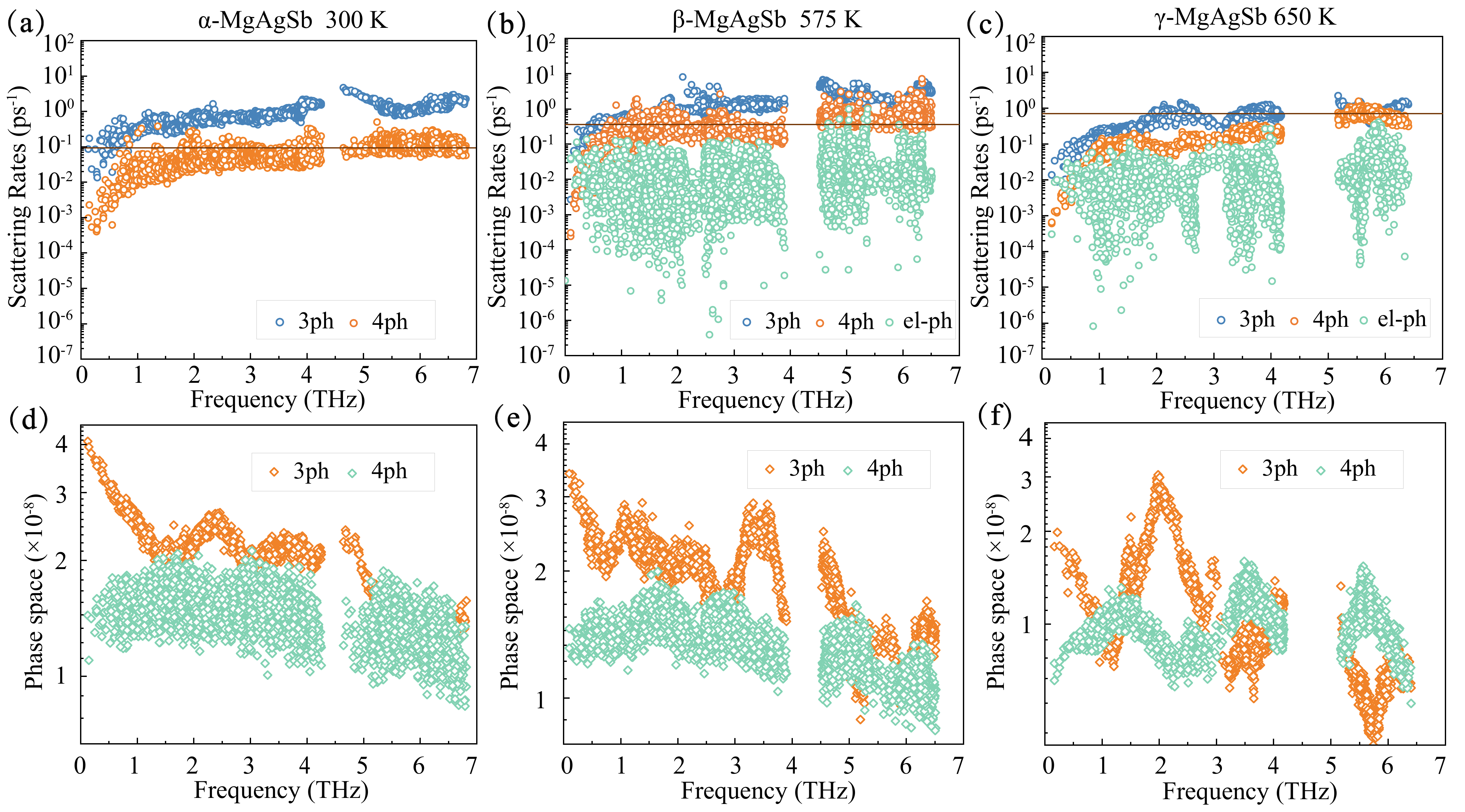, width=1\textwidth}
\caption{The phonon scattering rates of (a) $\alpha$ , (b) $\beta$, and (c) $\gamma$ at 300 K, 575 K, 650 K, respectively. The three-phonon ($P_3$) and four-phonon ($P_4$) phase space for (d) $\alpha$ , (e) $\beta$, and (f) $\gamma$ at 300 K, 575 K, 650 K, respectively.}
\label{Fig3}
\end{figure*}

Figure 3 presents the phonon scattering rates as a function of frequency. For the $\alpha$ phase at 300 K (Figure 3a), 3ph scattering dominates across the entire frequency range, with rates substantially higher than those of 4ph scattering. At low frequencies near 1~THz, the 3ph scattering rate is 0.31~ps$^{-1}$, whereas the 4ph contribution remains as low as 0.01~ps$^{-1}$. This negligible 4ph scattering is consistent with the minimal reduction in $\kappa_p$ observed when including 4ph processes in the $\alpha$ phase. In contrast, the $\beta$ at 575 K (Figure 3b) and $\gamma$ at 650 K (Figure 3c) phases exhibit more complex scattering characteristics. For the $\beta$ phase, 3ph scattering remains the strongest overall, followed by 4ph and el-ph scattering. Notably, in the frequency range of 1--2.5~THz, the 4ph scattering rates become comparable to those of 3ph processes, both lying within the same order of magnitude. Additionally, at very low frequencies (0--1~THz), el-ph scattering has a significant effect on the partial phonon modes, with rates approaching those of 3ph and 4ph scattering. Similarly, for the $\gamma$ phase, 3ph scattering generally dominates, yet 4ph scattering rates are comparable to 3ph in the 1--2~THz and 3--4~THz ranges. As in the $\beta$ phase, el-ph scattering also becomes appreciable for low-frequency modes below 1~THz. The enhanced role of 4ph scattering in the $\beta$ and $\gamma$ phases correlates directly with the notable reduction in $\kappa_p$ upon inclusion of 4ph processes. A horizontal dashed line in each panel marks the Wigner limit, defined as $\Delta\omega = \omega_{\text{max}} / 3n$, where $\omega_{\text{max}}$ is the maximum phonon frequency and $n$ is the number of atoms per primitive cell. Phonon modes above this line are considered to contribute predominantly to the $\kappa_c$\cite{DiLucente2023}. Compared to the $\alpha$ phase, the $\beta$ and $\gamma$ phases exhibit substantially fewer phonon modes above the Wigner limit, which is consistent with their lower $\kappa_c$ values, and further supports the explanation that in the more symmetric phase state, coherent phonon transmission is significantly suppressed.

To clarify how scattering rates influence $\kappa_L$ across different phases, Figure S1 compares the phonon scattering rates between different phases at the same temperature. All the three phases take into account all the scattering situations. At 575 K, the scattering rates of the $\alpha$ and $\beta$ phases are close (Figure S1a). Given that the total heat capacity ($c_v$) of the $\beta$ phase is only 0.9\% higher than that of the $\alpha$ phase, the lower $\kappa_L$ of the $\alpha$ phase can be primarily attributed to its reduced acoustic branch $c_v$ and lower $v_g$. At 650 K, the scattering rate of the $\beta$ phase is higher than that of the $\gamma$ phase (Figure S1b). For instance, near 1.1 THz, the scattering rate of the $\beta$ phase is  $\sim$0.89\,ps$^{-1}$, compared to only 0.21\,ps$^{-1}$ for the $\gamma$ phase. Additionally, the total $c_v$ of the $\beta$ phase is 4\% higher than that of the $\gamma$ phase. However, the $\beta$ phase larger unit cell size reduces the contribution of the acoustic branch, and the acoustic branch plays a dominant role in heat transfer. Therefore, phonon scattering is the main factor determining the difference in $\kappa_L$ between the $\beta$ and $\gamma$ phases.

\begin{figure*}[ht!]
\centering
\epsfig{file=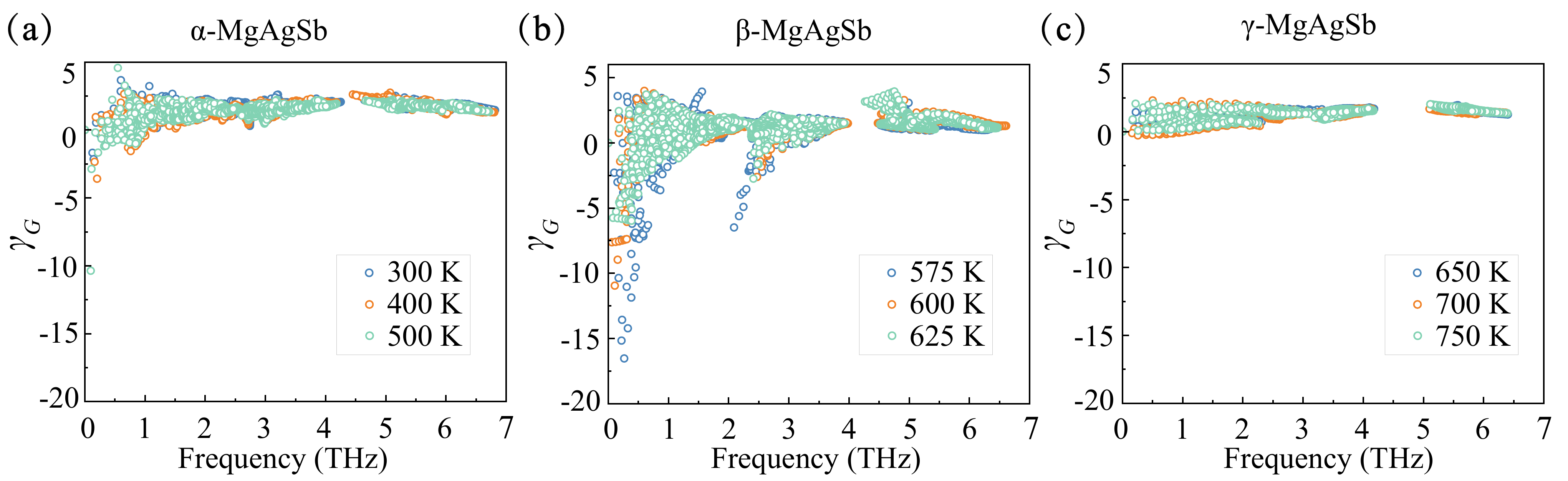, width=1\textwidth}
\caption{The Gr\"uneisen parameters ($\gamma_G$) of (a) $\alpha$-MgAgSb, (b) $\beta$-MgAgSb, and (c) $\gamma$-MgAgSb, at their respective temperatures.}
\label{Fig4}
\end{figure*}

The three-phonon ($P_3$) and four-phonon ($P_4$) phase space for the $\alpha$ phase at 300 K, the $\beta$ phase at 575 K, and the $\gamma$ phase at 650 K are presented in Figures 3(d)–(f). The larger phase space typically implies the existence of more available scattering channels, which simultaneously satisfy the conditions of energy and momentum conservation\cite{Li2024}. The $\alpha$ phase has a dense and flat phonon spectrum, which is conducive to satisfying the conservation conditions and makes it have the largest $P_3$ among the three phases. As the number of primitive cell atoms decreases from the $\alpha$ phase to the $\beta$ phase and then to the $\gamma$ phase, the phonon spectrum becomes sparse, resulting in a significant decrease in $P_3$. Meanwhile, $P_4$ also decreases with the simplification of the structure, but the decrease is smaller. Therefore, in the $\gamma$ phase, the contribution of $P_4$ becomes more significant.

Figure~4 shows the Gr\"uneisen parameters ($\gamma_G$) of the three MgAgSb phases at their respective temperatures, which reflect the anharmonic characteristics. Here, the degree of anharmonicity is evaluated by the absolute value of $\gamma_G$. In the $\alpha$ and $\gamma$ phases (Figures 4a and 4c), $|\gamma_G|$ does not change significantly within their respective temperature ranges. In contrast, for the $\beta$ phase (Figure 4b), the $|\gamma_G|$ decreases markedly with rising temperature, indicating a gradual weakening of lattice anharmonicity. This reduces the temperature dependence of $\kappa_p$ and leads to a slight increase in $\kappa_p$ between 575 and 650 K.

\section*{\Rmnum{4}.CONCLUSIONS}
In summary, we have systematically investigated the lattice thermal transport across the $\alpha$, $\beta$, and $\gamma$ phases of MgAgSb using first-principles calculations. Our results reveal a progressive increase in $\kappa_L$ following $\alpha < \beta < \gamma$, originating from distinct phase-dependent scattering characteristics. The 4ph scattering plays a crucial role in suppressing $\kappa_p$ in the $\beta$ and $\gamma$ phases, reducing it by 22.8\% at 575 K and 24.2\% at 650 K, respectively. While el-ph scattering provides additional reduction of 6.7\% and 9.2\%, respectively. The $\kappa_c$ is most significant in the structurally complex $\alpha$ phase, accounting for up to 44\% of the $\kappa_L$, and exhibits a pronounced increase with temperature. Meanwhile, the temperature dependence of $\kappa_p$ is weak, leading to a weak overall temperature dependence ($\kappa_L \sim T^{-0.33}$) of the $\alpha$ phase. The weak temperature dependence of the $\beta$ phase is attributed to a decreasing $\gamma_G$ with rising temperature. These findings establish a comprehensive microscopic picture of phase-dependent thermal transport in MgAgSb.

\section*{\Rmnum{4}.Acknowledgements}
This work is supported by the Natural Science Foundation of China (Grants No. 12304038, 52206092, 12204402), the National Key Research and Development Program of China (Grants No. 2024YFA1409800), the Big Data Computing Center of Southeast University and the Center for Fundamental and Interdisciplinary Sciences of Southeast University, Basic Research Program of Jiangsu (BK20250035), Major Basic Research Project of the Natural Science Foundation of the Jiangsu Higher Education Institutions (25KJA470006).



%


\begin{thebibliography}{60}%
\makeatletter
\providecommand \@ifxundefined [1]{%
 \@ifx{#1\undefined}
}%
\providecommand \@ifnum [1]{%
 \ifnum #1\expandafter \@firstoftwo
 \else \expandafter \@secondoftwo
 \fi
}%
\providecommand \@ifx [1]{%
 \ifx #1\expandafter \@firstoftwo
 \else \expandafter \@secondoftwo
 \fi
}%
\providecommand \natexlab [1]{#1}%
\providecommand \enquote  [1]{``#1''}%
\providecommand \bibnamefont  [1]{#1}%
\providecommand \bibfnamefont [1]{#1}%
\providecommand \citenamefont [1]{#1}%
\providecommand \href@noop [0]{\@secondoftwo}%
\providecommand \href [0]{\begingroup \@sanitize@url \@href}%
\providecommand \@href[1]{\@@startlink{#1}\@@href}%
\providecommand \@@href[1]{\endgroup#1\@@endlink}%
\providecommand \@sanitize@url [0]{\catcode `\\12\catcode `\$12\catcode
  `\&12\catcode `\#12\catcode `\^12\catcode `\_12\catcode `\%12\relax}%
\providecommand \@@startlink[1]{}%
\providecommand \@@endlink[0]{}%
\providecommand \url  [0]{\begingroup\@sanitize@url \@url }%
\providecommand \@url [1]{\endgroup\@href {#1}{\urlprefix }}%
\providecommand \urlprefix  [0]{URL }%
\providecommand \Eprint [0]{\href }%
\providecommand \doibase [0]{https://doi.org/}%
\providecommand \selectlanguage [0]{\@gobble}%
\providecommand \bibinfo  [0]{\@secondoftwo}%
\providecommand \bibfield  [0]{\@secondoftwo}%
\providecommand \translation [1]{[#1]}%
\providecommand \BibitemOpen [0]{}%
\providecommand \bibitemStop [0]{}%
\providecommand \bibitemNoStop [0]{.\EOS\space}%
\providecommand \EOS [0]{\spacefactor3000\relax}%
\providecommand \BibitemShut  [1]{\csname bibitem#1\endcsname}%
\let\auto@bib@innerbib\@empty
\bibitem [{\citenamefont {Wang}\ \emph
  {et~al.}(2023{\natexlab{a}})\citenamefont {Wang}, \citenamefont {Tang},
  \citenamefont {Gao}, \citenamefont {Liu}, \citenamefont {Li}, \citenamefont
  {Li},\ and\ \citenamefont {Chen}}]{Wang2023c}%
  \BibitemOpen
  \bibfield  {author} {\bibinfo {author} {\bibfnamefont {G.}~\bibnamefont
  {Wang}}, \bibinfo {author} {\bibfnamefont {Z.}~\bibnamefont {Tang}}, \bibinfo
  {author} {\bibfnamefont {Y.}~\bibnamefont {Gao}}, \bibinfo {author}
  {\bibfnamefont {P.}~\bibnamefont {Liu}}, \bibinfo {author} {\bibfnamefont
  {Y.}~\bibnamefont {Li}}, \bibinfo {author} {\bibfnamefont {A.}~\bibnamefont
  {Li}},\ and\ \bibinfo {author} {\bibfnamefont {X.}~\bibnamefont {Chen}},\
  }\bibfield  {title} {\bibinfo {title} {Phase change thermal storage materials
  for interdisciplinary applications},\ }\href
  {https://doi.org/10.1021/acs.chemrev.2c00572} {\bibfield  {journal} {\bibinfo
   {journal} {Chem. Rev.}\ }\textbf {\bibinfo {volume} {123}},\ \bibinfo
  {pages} {6953} (\bibinfo {year} {2023}{\natexlab{a}})}\BibitemShut {NoStop}%
\bibitem [{\citenamefont {Jiang}\ \emph {et~al.}(2018)\citenamefont {Jiang},
  \citenamefont {Qiu}, \citenamefont {Chen}, \citenamefont {Huang},
  \citenamefont {Mao}, \citenamefont {Wang}, \citenamefont {Song},
  \citenamefont {Ren}, \citenamefont {Shi},\ and\ \citenamefont
  {Chen}}]{Jiang2018a}%
  \BibitemOpen
  \bibfield  {author} {\bibinfo {author} {\bibfnamefont {B.}~\bibnamefont
  {Jiang}}, \bibinfo {author} {\bibfnamefont {P.}~\bibnamefont {Qiu}}, \bibinfo
  {author} {\bibfnamefont {H.}~\bibnamefont {Chen}}, \bibinfo {author}
  {\bibfnamefont {J.}~\bibnamefont {Huang}}, \bibinfo {author} {\bibfnamefont
  {T.}~\bibnamefont {Mao}}, \bibinfo {author} {\bibfnamefont {Y.}~\bibnamefont
  {Wang}}, \bibinfo {author} {\bibfnamefont {Q.}~\bibnamefont {Song}}, \bibinfo
  {author} {\bibfnamefont {D.}~\bibnamefont {Ren}}, \bibinfo {author}
  {\bibfnamefont {X.}~\bibnamefont {Shi}},\ and\ \bibinfo {author}
  {\bibfnamefont {L.}~\bibnamefont {Chen}},\ }\bibfield  {title} {\bibinfo
  {title} {Entropy optimized phase transitions and improved thermoelectric
  performance in n-type liquid-like {Ag$_9$GaSe$_6$} materials},\ }\href
  {https://doi.org/10.1016/j.mtphys.2018.05.001} {\bibfield  {journal}
  {\bibinfo  {journal} {Mater. Today Phys.}\ }\textbf {\bibinfo {volume} {5}},\
  \bibinfo {pages} {20} (\bibinfo {year} {2018})}\BibitemShut {NoStop}%
\bibitem [{\citenamefont {Yu}\ and\ \citenamefont {Wuttig}(2025)}]{Yu2025}%
  \BibitemOpen
  \bibfield  {author} {\bibinfo {author} {\bibfnamefont {Y.}~\bibnamefont
  {Yu}}\ and\ \bibinfo {author} {\bibfnamefont {M.}~\bibnamefont {Wuttig}},\
  }\bibfield  {title} {\bibinfo {title} {From phase-change materials to
  thermoelectrics: The role of metavalent bonding},\ }\href
  {https://doi.org/10.1557/s43578-025-01619-2} {\bibfield  {journal} {\bibinfo
  {journal} {J. Mater. Res.}\ }\textbf {\bibinfo {volume} {40}},\ \bibinfo
  {pages} {1747} (\bibinfo {year} {2025})}\BibitemShut {NoStop}%
\bibitem [{\citenamefont {Shi}\ \emph {et~al.}(2020)\citenamefont {Shi},
  \citenamefont {Zou},\ and\ \citenamefont {Chen}}]{Shi2020}%
  \BibitemOpen
  \bibfield  {author} {\bibinfo {author} {\bibfnamefont {X.-L.}\ \bibnamefont
  {Shi}}, \bibinfo {author} {\bibfnamefont {J.}~\bibnamefont {Zou}},\ and\
  \bibinfo {author} {\bibfnamefont {Z.-G.}\ \bibnamefont {Chen}},\ }\bibfield
  {title} {\bibinfo {title} {Advanced thermoelectric design: From materials and
  structures to devices},\ }\href {https://doi.org/10.1021/acs.chemrev.0c00026}
  {\bibfield  {journal} {\bibinfo  {journal} {Chem. Rev.}\ }\textbf {\bibinfo
  {volume} {120}},\ \bibinfo {pages} {7399} (\bibinfo {year}
  {2020})}\BibitemShut {NoStop}%
\bibitem [{\citenamefont {Chebli}\ and\ \citenamefont
  {Mechighel}(2025)}]{Chebli2025}%
  \BibitemOpen
  \bibfield  {author} {\bibinfo {author} {\bibfnamefont {F.}~\bibnamefont
  {Chebli}}\ and\ \bibinfo {author} {\bibfnamefont {F.}~\bibnamefont
  {Mechighel}},\ }\bibfield  {title} {\bibinfo {title} {Phase change materials:
  classification, use, phase transitions, and heat transfer enhancement
  techniques: a comprehensive review},\ }\href
  {https://doi.org/10.1007/s10973-024-13877-z} {\bibfield  {journal} {\bibinfo
  {journal} {J. Therm. Anal. Calorim.}\ }\textbf {\bibinfo {volume} {150}},\
  \bibinfo {pages} {1353} (\bibinfo {year} {2025})}\BibitemShut {NoStop}%
\bibitem [{\citenamefont {Omini}\ and\ \citenamefont
  {Sparavigna}(1995)}]{Omini1995}%
  \BibitemOpen
  \bibfield  {author} {\bibinfo {author} {\bibfnamefont {M.}~\bibnamefont
  {Omini}}\ and\ \bibinfo {author} {\bibfnamefont {A.}~\bibnamefont
  {Sparavigna}},\ }\bibfield  {title} {\bibinfo {title} {An iterative approach
  to the phonon boltzmann equation in the theory of thermal conductivity},\
  }\href {https://doi.org/10.1016/0921-4526(95)00016-3} {\bibfield  {journal}
  {\bibinfo  {journal} {Physica B.}\ }\textbf {\bibinfo {volume} {212}},\
  \bibinfo {pages} {101} (\bibinfo {year} {1995})}\BibitemShut {NoStop}%
\bibitem [{\citenamefont {Yang}\ \emph {et~al.}(2018)\citenamefont {Yang},
  \citenamefont {Dai}, \citenamefont {Zhao}, \citenamefont {Liu},\ and\
  \citenamefont {Meng}}]{Yang2018}%
  \BibitemOpen
  \bibfield  {author} {\bibinfo {author} {\bibfnamefont {X.}~\bibnamefont
  {Yang}}, \bibinfo {author} {\bibfnamefont {Z.}~\bibnamefont {Dai}}, \bibinfo
  {author} {\bibfnamefont {Y.}~\bibnamefont {Zhao}}, \bibinfo {author}
  {\bibfnamefont {J.}~\bibnamefont {Liu}},\ and\ \bibinfo {author}
  {\bibfnamefont {S.}~\bibnamefont {Meng}},\ }\bibfield  {title} {\bibinfo
  {title} {Low lattice thermal conductivity and excellent thermoelectric
  behavior in {Li$_3$Sb} and {Li$_3$Bi}},\ }\href
  {https://doi.org/10.1088/1361-648x/aade17} {\bibfield  {journal} {\bibinfo
  {journal} {J. Phys.:Condens. Mat.}\ }\textbf {\bibinfo {volume} {30}},\
  \bibinfo {pages} {425401} (\bibinfo {year} {2018})}\BibitemShut {NoStop}%
\bibitem [{\citenamefont {Li}\ \emph {et~al.}(2021)\citenamefont {Li},
  \citenamefont {Xie}, \citenamefont {Hao}, \citenamefont {Xia}, \citenamefont
  {Su}, \citenamefont {Kanatzidis}, \citenamefont {Wolverton},\ and\
  \citenamefont {Tang}}]{Li2021}%
  \BibitemOpen
  \bibfield  {author} {\bibinfo {author} {\bibfnamefont {Z.}~\bibnamefont
  {Li}}, \bibinfo {author} {\bibfnamefont {H.}~\bibnamefont {Xie}}, \bibinfo
  {author} {\bibfnamefont {S.}~\bibnamefont {Hao}}, \bibinfo {author}
  {\bibfnamefont {Y.}~\bibnamefont {Xia}}, \bibinfo {author} {\bibfnamefont
  {X.}~\bibnamefont {Su}}, \bibinfo {author} {\bibfnamefont {M.~G.}\
  \bibnamefont {Kanatzidis}}, \bibinfo {author} {\bibfnamefont
  {C.}~\bibnamefont {Wolverton}},\ and\ \bibinfo {author} {\bibfnamefont
  {X.}~\bibnamefont {Tang}},\ }\bibfield  {title} {\bibinfo {title} {Optical
  phonon dominated heat transport: A first-principles thermal conductivity
  study of {BaSnS$_2$}},\ }\href {https://doi.org/10.1103/physrevb.104.245209}
  {\bibfield  {journal} {\bibinfo  {journal} {Phys. Rev. B}\ }\textbf {\bibinfo
  {volume} {104}},\ \bibinfo {pages} {245209} (\bibinfo {year}
  {2021})}\BibitemShut {NoStop}%
\bibitem [{\citenamefont {Luckyanova}\ \emph {et~al.}(2012)\citenamefont
  {Luckyanova}, \citenamefont {Garg}, \citenamefont {Esfarjani}, \citenamefont
  {Jandl}, \citenamefont {Bulsara}, \citenamefont {Schmidt}, \citenamefont
  {Minnich}, \citenamefont {Chen}, \citenamefont {Dresselhaus}, \citenamefont
  {Ren}, \citenamefont {Fitzgerald},\ and\ \citenamefont
  {Chen}}]{Luckyanova2012}%
  \BibitemOpen
  \bibfield  {author} {\bibinfo {author} {\bibfnamefont {M.~N.}\ \bibnamefont
  {Luckyanova}}, \bibinfo {author} {\bibfnamefont {J.}~\bibnamefont {Garg}},
  \bibinfo {author} {\bibfnamefont {K.}~\bibnamefont {Esfarjani}}, \bibinfo
  {author} {\bibfnamefont {A.}~\bibnamefont {Jandl}}, \bibinfo {author}
  {\bibfnamefont {M.~T.}\ \bibnamefont {Bulsara}}, \bibinfo {author}
  {\bibfnamefont {A.~J.}\ \bibnamefont {Schmidt}}, \bibinfo {author}
  {\bibfnamefont {A.~J.}\ \bibnamefont {Minnich}}, \bibinfo {author}
  {\bibfnamefont {S.}~\bibnamefont {Chen}}, \bibinfo {author} {\bibfnamefont
  {M.~S.}\ \bibnamefont {Dresselhaus}}, \bibinfo {author} {\bibfnamefont
  {Z.}~\bibnamefont {Ren}}, \bibinfo {author} {\bibfnamefont {E.~A.}\
  \bibnamefont {Fitzgerald}},\ and\ \bibinfo {author} {\bibfnamefont
  {G.}~\bibnamefont {Chen}},\ }\bibfield  {title} {\bibinfo {title} {Coherent
  phonon heat conduction in superlattices},\ }\href
  {https://doi.org/10.1126/science.1225549} {\bibfield  {journal} {\bibinfo
  {journal} {Science}\ }\textbf {\bibinfo {volume} {338}},\ \bibinfo {pages}
  {936} (\bibinfo {year} {2012})}\BibitemShut {NoStop}%
\bibitem [{\citenamefont {Wang}\ \emph {et~al.}(2005)\citenamefont {Wang},
  \citenamefont {Hashimoto}, \citenamefont {Kono}, \citenamefont {Oiwa},
  \citenamefont {Munekata}, \citenamefont {Sanders},\ and\ \citenamefont
  {Stanton}}]{Wang2005}%
  \BibitemOpen
  \bibfield  {author} {\bibinfo {author} {\bibfnamefont {J.}~\bibnamefont
  {Wang}}, \bibinfo {author} {\bibfnamefont {Y.}~\bibnamefont {Hashimoto}},
  \bibinfo {author} {\bibfnamefont {J.}~\bibnamefont {Kono}}, \bibinfo {author}
  {\bibfnamefont {A.}~\bibnamefont {Oiwa}}, \bibinfo {author} {\bibfnamefont
  {H.}~\bibnamefont {Munekata}}, \bibinfo {author} {\bibfnamefont {G.~D.}\
  \bibnamefont {Sanders}},\ and\ \bibinfo {author} {\bibfnamefont {C.~J.}\
  \bibnamefont {Stanton}},\ }\bibfield  {title} {\bibinfo {title} {Propagating
  coherent acoustic phonon wave packets in {In$_x$Mn$_{1-x}$As/GaSb}},\ }\href
  {https://doi.org/10.1103/physrevb.72.153311} {\bibfield  {journal} {\bibinfo
  {journal} {Phys. Rev. B}\ }\textbf {\bibinfo {volume} {72}},\ \bibinfo
  {pages} {153311} (\bibinfo {year} {2005})}\BibitemShut {NoStop}%
\bibitem [{\citenamefont {Colvard}\ \emph {et~al.}(1980)\citenamefont
  {Colvard}, \citenamefont {Merlin}, \citenamefont {Klein},\ and\ \citenamefont
  {Gossard}}]{Colvard1980}%
  \BibitemOpen
  \bibfield  {author} {\bibinfo {author} {\bibfnamefont {C.}~\bibnamefont
  {Colvard}}, \bibinfo {author} {\bibfnamefont {R.}~\bibnamefont {Merlin}},
  \bibinfo {author} {\bibfnamefont {M.~V.}\ \bibnamefont {Klein}},\ and\
  \bibinfo {author} {\bibfnamefont {A.~C.}\ \bibnamefont {Gossard}},\
  }\bibfield  {title} {\bibinfo {title} {Observation of folded acoustic phonons
  in a semiconductor superlattice},\ }\href
  {https://doi.org/10.1103/physrevlett.45.298} {\bibfield  {journal} {\bibinfo
  {journal} {Phys. Rev. Lett.}\ }\textbf {\bibinfo {volume} {45}},\ \bibinfo
  {pages} {298} (\bibinfo {year} {1980})}\BibitemShut {NoStop}%
\bibitem [{\citenamefont {Zeng}\ \emph {et~al.}(2025)\citenamefont {Zeng},
  \citenamefont {Niu}, \citenamefont {Wang}, \citenamefont {Jiang},
  \citenamefont {Sui}, \citenamefont {Zhang}, \citenamefont {Chen},
  \citenamefont {Jin},\ and\ \citenamefont {Yuan}}]{Zeng2025}%
  \BibitemOpen
  \bibfield  {author} {\bibinfo {author} {\bibfnamefont {X.}~\bibnamefont
  {Zeng}}, \bibinfo {author} {\bibfnamefont {G.}~\bibnamefont {Niu}}, \bibinfo
  {author} {\bibfnamefont {X.}~\bibnamefont {Wang}}, \bibinfo {author}
  {\bibfnamefont {J.}~\bibnamefont {Jiang}}, \bibinfo {author} {\bibfnamefont
  {L.}~\bibnamefont {Sui}}, \bibinfo {author} {\bibfnamefont {Y.}~\bibnamefont
  {Zhang}}, \bibinfo {author} {\bibfnamefont {A.}~\bibnamefont {Chen}},
  \bibinfo {author} {\bibfnamefont {M.}~\bibnamefont {Jin}},\ and\ \bibinfo
  {author} {\bibfnamefont {K.}~\bibnamefont {Yuan}},\ }\bibfield  {title}
  {\bibinfo {title} {Strain-engineered thermoelectric performance in
  superatomic semiconductor {Re$_6$Se$_8$Cl$_2$}: The role of four-phonon
  scattering and coherent phonons},\ }\href
  {https://doi.org/10.1021/acs.jpclett.5c00404} {\bibfield  {journal} {\bibinfo
   {journal} {J. Phys. Chem. Lett.}\ }\textbf {\bibinfo {volume} {16}},\
  \bibinfo {pages} {3122} (\bibinfo {year} {2025})}\BibitemShut {NoStop}%
\bibitem [{\citenamefont {Simoncelli}\ \emph {et~al.}(2022)\citenamefont
  {Simoncelli}, \citenamefont {Marzari},\ and\ \citenamefont
  {Mauri}}]{Simoncelli2022}%
  \BibitemOpen
  \bibfield  {author} {\bibinfo {author} {\bibfnamefont {M.}~\bibnamefont
  {Simoncelli}}, \bibinfo {author} {\bibfnamefont {N.}~\bibnamefont
  {Marzari}},\ and\ \bibinfo {author} {\bibfnamefont {F.}~\bibnamefont
  {Mauri}},\ }\bibfield  {title} {\bibinfo {title} {Wigner formulation of
  thermal transport in solids},\ }\href
  {https://doi.org/10.1103/physrevx.12.041011} {\bibfield  {journal} {\bibinfo
  {journal} {Phys. Rev. X}\ }\textbf {\bibinfo {volume} {12}},\ \bibinfo
  {pages} {041011} (\bibinfo {year} {2022})}\BibitemShut {NoStop}%
\bibitem [{\citenamefont {Legenstein}\ \emph {et~al.}(2025)\citenamefont
  {Legenstein}, \citenamefont {Reicht}, \citenamefont {Wieser}, \citenamefont
  {Simoncelli},\ and\ \citenamefont {Zojer}}]{Legenstein2025}%
  \BibitemOpen
  \bibfield  {author} {\bibinfo {author} {\bibfnamefont {L.}~\bibnamefont
  {Legenstein}}, \bibinfo {author} {\bibfnamefont {L.}~\bibnamefont {Reicht}},
  \bibinfo {author} {\bibfnamefont {S.}~\bibnamefont {Wieser}}, \bibinfo
  {author} {\bibfnamefont {M.}~\bibnamefont {Simoncelli}},\ and\ \bibinfo
  {author} {\bibfnamefont {E.}~\bibnamefont {Zojer}},\ }\bibfield  {title}
  {\bibinfo {title} {Heat transport in crystalline organic semiconductors:
  coexistence of phonon propagation and tunneling},\ }\href
  {https://doi.org/10.1038/s41524-025-01514-8} {\bibfield  {journal} {\bibinfo
  {journal} {npj Comput. Mater.}\ }\textbf {\bibinfo {volume} {11}},\ \bibinfo
  {pages} {29} (\bibinfo {year} {2025})}\BibitemShut {NoStop}%
\bibitem [{\citenamefont {Isaeva}\ \emph {et~al.}(2019)\citenamefont {Isaeva},
  \citenamefont {Barbalinardo}, \citenamefont {Donadio},\ and\ \citenamefont
  {Baroni}}]{Isaeva2019}%
  \BibitemOpen
  \bibfield  {author} {\bibinfo {author} {\bibfnamefont {L.}~\bibnamefont
  {Isaeva}}, \bibinfo {author} {\bibfnamefont {G.}~\bibnamefont
  {Barbalinardo}}, \bibinfo {author} {\bibfnamefont {D.}~\bibnamefont
  {Donadio}},\ and\ \bibinfo {author} {\bibfnamefont {S.}~\bibnamefont
  {Baroni}},\ }\bibfield  {title} {\bibinfo {title} {Modeling heat transport in
  crystals and glasses from a unified lattice-dynamical approach},\ }\href
  {https://doi.org/10.1038/s41467-019-11572-4} {\bibfield  {journal} {\bibinfo
  {journal} {Nat. Commu.}\ }\textbf {\bibinfo {volume} {10}},\ \bibinfo {pages}
  {3853} (\bibinfo {year} {2019})}\BibitemShut {NoStop}%
\bibitem [{\citenamefont {Mukhopadhyay}\ \emph {et~al.}(2018)\citenamefont
  {Mukhopadhyay}, \citenamefont {Parker}, \citenamefont {Sales}, \citenamefont
  {Puretzky}, \citenamefont {McGuire},\ and\ \citenamefont
  {Lindsay}}]{Mukhopadhyay2018}%
  \BibitemOpen
  \bibfield  {author} {\bibinfo {author} {\bibfnamefont {S.}~\bibnamefont
  {Mukhopadhyay}}, \bibinfo {author} {\bibfnamefont {D.~S.}\ \bibnamefont
  {Parker}}, \bibinfo {author} {\bibfnamefont {B.~C.}\ \bibnamefont {Sales}},
  \bibinfo {author} {\bibfnamefont {A.~A.}\ \bibnamefont {Puretzky}}, \bibinfo
  {author} {\bibfnamefont {M.~A.}\ \bibnamefont {McGuire}},\ and\ \bibinfo
  {author} {\bibfnamefont {L.}~\bibnamefont {Lindsay}},\ }\bibfield  {title}
  {\bibinfo {title} {Two-channel model for ultralow thermal conductivity of
  crystalline {Tl$_3$VSe$_4$}},\ }\href
  {https://doi.org/10.1126/science.aar8072} {\bibfield  {journal} {\bibinfo
  {journal} {Science}\ }\textbf {\bibinfo {volume} {360}},\ \bibinfo {pages}
  {1455} (\bibinfo {year} {2018})}\BibitemShut {NoStop}%
\bibitem [{\citenamefont {Shenogin}\ \emph {et~al.}(2009)\citenamefont
  {Shenogin}, \citenamefont {Bodapati}, \citenamefont {Keblinski},\ and\
  \citenamefont {McGaughey}}]{Shenogin2009}%
  \BibitemOpen
  \bibfield  {author} {\bibinfo {author} {\bibfnamefont {S.}~\bibnamefont
  {Shenogin}}, \bibinfo {author} {\bibfnamefont {A.}~\bibnamefont {Bodapati}},
  \bibinfo {author} {\bibfnamefont {P.}~\bibnamefont {Keblinski}},\ and\
  \bibinfo {author} {\bibfnamefont {A.~J.~H.}\ \bibnamefont {McGaughey}},\
  }\bibfield  {title} {\bibinfo {title} {Predicting the thermal conductivity of
  inorganic and polymeric glasses: The role of anharmonicity},\ }\href
  {https://doi.org/10.1063/1.3073954} {\bibfield  {journal} {\bibinfo
  {journal} {J. Appl. Phys.}\ }\textbf {\bibinfo {volume} {105}},\ \bibinfo
  {pages} {034906} (\bibinfo {year} {2009})}\BibitemShut {NoStop}%
\bibitem [{\citenamefont {Zhang}\ \emph {et~al.}(2021)\citenamefont {Zhang},
  \citenamefont {Guo}, \citenamefont {Bescond}, \citenamefont {Chen},
  \citenamefont {Nomura},\ and\ \citenamefont {Volz}}]{Zhang2021}%
  \BibitemOpen
  \bibfield  {author} {\bibinfo {author} {\bibfnamefont {Z.}~\bibnamefont
  {Zhang}}, \bibinfo {author} {\bibfnamefont {Y.}~\bibnamefont {Guo}}, \bibinfo
  {author} {\bibfnamefont {M.}~\bibnamefont {Bescond}}, \bibinfo {author}
  {\bibfnamefont {J.}~\bibnamefont {Chen}}, \bibinfo {author} {\bibfnamefont
  {M.}~\bibnamefont {Nomura}},\ and\ \bibinfo {author} {\bibfnamefont
  {S.}~\bibnamefont {Volz}},\ }\bibfield  {title} {\bibinfo {title}
  {Generalized decay law for particlelike and wavelike thermal phonons},\
  }\href {https://doi.org/10.1103/physrevb.103.184307} {\bibfield  {journal}
  {\bibinfo  {journal} {Phys. Rev. B}\ }\textbf {\bibinfo {volume} {103}},\
  \bibinfo {pages} {184307} (\bibinfo {year} {2021})}\BibitemShut {NoStop}%
\bibitem [{\citenamefont {Das}\ and\ \citenamefont {Banerji}(2023)}]{Das2023}%
  \BibitemOpen
  \bibfield  {author} {\bibinfo {author} {\bibfnamefont {A.}~\bibnamefont
  {Das}}\ and\ \bibinfo {author} {\bibfnamefont {P.}~\bibnamefont {Banerji}},\
  }\bibfield  {title} {\bibinfo {title} {Antibonding or nonbonding
  interaction-driven phonon modes softening and wave-like interband thermal
  conduction in layered {In$_4$Te$_3$} under the framework of wigner transport
  formalism},\ }\href {https://doi.org/10.1021/acsaem.3c01838} {\bibfield
  {journal} {\bibinfo  {journal} {ACS Appl. Energy Mater.}\ }\textbf {\bibinfo
  {volume} {6}},\ \bibinfo {pages} {11521} (\bibinfo {year}
  {2023})}\BibitemShut {NoStop}%
\bibitem [{\citenamefont {Yang}\ \emph {et~al.}(2019)\citenamefont {Yang},
  \citenamefont {Feng}, \citenamefont {Li},\ and\ \citenamefont
  {Ruan}}]{Yang2019}%
  \BibitemOpen
  \bibfield  {author} {\bibinfo {author} {\bibfnamefont {X.}~\bibnamefont
  {Yang}}, \bibinfo {author} {\bibfnamefont {T.}~\bibnamefont {Feng}}, \bibinfo
  {author} {\bibfnamefont {J.}~\bibnamefont {Li}},\ and\ \bibinfo {author}
  {\bibfnamefont {X.}~\bibnamefont {Ruan}},\ }\bibfield  {title} {\bibinfo
  {title} {Stronger role of four-phonon scattering than three-phonon scattering
  in thermal conductivity of {III-V} semiconductors at room temperature},\
  }\href {https://doi.org/10.1103/physrevb.100.245203} {\bibfield  {journal}
  {\bibinfo  {journal} {Phys. Rev. B}\ }\textbf {\bibinfo {volume} {100}},\
  \bibinfo {pages} {245203} (\bibinfo {year} {2019})}\BibitemShut {NoStop}%
\bibitem [{\citenamefont {Xia}\ \emph {et~al.}(2020)\citenamefont {Xia},
  \citenamefont {Hegde}, \citenamefont {Pal}, \citenamefont {Hua},
  \citenamefont {Gaines}, \citenamefont {Patel}, \citenamefont {He},
  \citenamefont {Aykol},\ and\ \citenamefont {Wolverton}}]{Xia2020}%
  \BibitemOpen
  \bibfield  {author} {\bibinfo {author} {\bibfnamefont {Y.}~\bibnamefont
  {Xia}}, \bibinfo {author} {\bibfnamefont {V.~I.}\ \bibnamefont {Hegde}},
  \bibinfo {author} {\bibfnamefont {K.}~\bibnamefont {Pal}}, \bibinfo {author}
  {\bibfnamefont {X.}~\bibnamefont {Hua}}, \bibinfo {author} {\bibfnamefont
  {D.}~\bibnamefont {Gaines}}, \bibinfo {author} {\bibfnamefont
  {S.}~\bibnamefont {Patel}}, \bibinfo {author} {\bibfnamefont
  {J.}~\bibnamefont {He}}, \bibinfo {author} {\bibfnamefont {M.}~\bibnamefont
  {Aykol}},\ and\ \bibinfo {author} {\bibfnamefont {C.}~\bibnamefont
  {Wolverton}},\ }\bibfield  {title} {\bibinfo {title} {High-throughput study
  of lattice thermal conductivity in binary rocksalt and zinc blende compounds
  including higher-order anharmonicity},\ }\href
  {https://doi.org/10.1103/physrevx.10.041029} {\bibfield  {journal} {\bibinfo
  {journal} {Phys. Rev. X}\ }\textbf {\bibinfo {volume} {10}},\ \bibinfo
  {pages} {041029} (\bibinfo {year} {2020})}\BibitemShut {NoStop}%
\bibitem [{\citenamefont {Feng}\ and\ \citenamefont {Ruan}(2018)}]{Feng2018}%
  \BibitemOpen
  \bibfield  {author} {\bibinfo {author} {\bibfnamefont {T.}~\bibnamefont
  {Feng}}\ and\ \bibinfo {author} {\bibfnamefont {X.}~\bibnamefont {Ruan}},\
  }\bibfield  {title} {\bibinfo {title} {Erratum: Quantum mechanical prediction
  of four-phonon scattering rates and reduced thermal conductivity of solids},\
  }\href {https://doi.org/10.1103/physrevb.97.079901} {\bibfield  {journal}
  {\bibinfo  {journal} {Phys. Rev. B}\ }\textbf {\bibinfo {volume} {97}},\
  \bibinfo {pages} {079901} (\bibinfo {year} {2018})}\BibitemShut {NoStop}%
\bibitem [{\citenamefont {Ravichandran}\ and\ \citenamefont
  {Broido}(2020)}]{Ravichandran2020}%
  \BibitemOpen
  \bibfield  {author} {\bibinfo {author} {\bibfnamefont {N.~K.}\ \bibnamefont
  {Ravichandran}}\ and\ \bibinfo {author} {\bibfnamefont {D.}~\bibnamefont
  {Broido}},\ }\bibfield  {title} {\bibinfo {title} {Phonon-phonon interactions
  in strongly bonded solids: Selection rules and higher-order processes},\
  }\href {https://doi.org/10.1103/physrevx.10.021063} {\bibfield  {journal}
  {\bibinfo  {journal} {Phys. Rev. X}\ }\textbf {\bibinfo {volume} {10}},\
  \bibinfo {pages} {021063} (\bibinfo {year} {2020})}\BibitemShut {NoStop}%
\bibitem [{\citenamefont {Ji}\ \emph {et~al.}(2024)\citenamefont {Ji},
  \citenamefont {Huang}, \citenamefont {Huo}, \citenamefont {Ding},
  \citenamefont {Zeng}, \citenamefont {Wu},\ and\ \citenamefont
  {Zhou}}]{Ji2024}%
  \BibitemOpen
  \bibfield  {author} {\bibinfo {author} {\bibfnamefont {L.}~\bibnamefont
  {Ji}}, \bibinfo {author} {\bibfnamefont {A.}~\bibnamefont {Huang}}, \bibinfo
  {author} {\bibfnamefont {Y.}~\bibnamefont {Huo}}, \bibinfo {author}
  {\bibfnamefont {Y.-m.}\ \bibnamefont {Ding}}, \bibinfo {author}
  {\bibfnamefont {S.}~\bibnamefont {Zeng}}, \bibinfo {author} {\bibfnamefont
  {Y.}~\bibnamefont {Wu}},\ and\ \bibinfo {author} {\bibfnamefont
  {L.}~\bibnamefont {Zhou}},\ }\bibfield  {title} {\bibinfo {title} {Influence
  of four-phonon scattering and wavelike phonon tunneling effects on the
  thermal transport properties of {TlBiSe$_2$}},\ }\href
  {https://doi.org/10.1103/physrevb.109.214307} {\bibfield  {journal} {\bibinfo
   {journal} {Phys. Rev. B}\ }\textbf {\bibinfo {volume} {109}},\ \bibinfo
  {pages} {214307} (\bibinfo {year} {2024})}\BibitemShut {NoStop}%
\bibitem [{\citenamefont {Tadano}\ and\ \citenamefont
  {Tsuneyuki}(2015)}]{Tadano2015a}%
  \BibitemOpen
  \bibfield  {author} {\bibinfo {author} {\bibfnamefont {T.}~\bibnamefont
  {Tadano}}\ and\ \bibinfo {author} {\bibfnamefont {S.}~\bibnamefont
  {Tsuneyuki}},\ }\bibfield  {title} {\bibinfo {title} {Self-consistent phonon
  calculations of lattice dynamical properties in cubic {SrTiO$_3$} with
  first-principles anharmonic force constants},\ }\href
  {https://doi.org/10.1103/physrevb.92.054301} {\bibfield  {journal} {\bibinfo
  {journal} {Phys. Rev. B}\ }\textbf {\bibinfo {volume} {92}},\ \bibinfo
  {pages} {054301} (\bibinfo {year} {2015})}\BibitemShut {NoStop}%
\bibitem [{\citenamefont {Simoncelli}\ \emph {et~al.}(2019)\citenamefont
  {Simoncelli}, \citenamefont {Marzari},\ and\ \citenamefont
  {Mauri}}]{Simoncelli2019}%
  \BibitemOpen
  \bibfield  {author} {\bibinfo {author} {\bibfnamefont {M.}~\bibnamefont
  {Simoncelli}}, \bibinfo {author} {\bibfnamefont {N.}~\bibnamefont
  {Marzari}},\ and\ \bibinfo {author} {\bibfnamefont {F.}~\bibnamefont
  {Mauri}},\ }\bibfield  {title} {\bibinfo {title} {Unified theory of thermal
  transport in crystals and glasses},\ }\href
  {https://doi.org/10.1038/s41567-019-0520-x} {\bibfield  {journal} {\bibinfo
  {journal} {Nat. Phys.}\ }\textbf {\bibinfo {volume} {15}},\ \bibinfo {pages}
  {809} (\bibinfo {year} {2019})}\BibitemShut {NoStop}%
\bibitem [{\citenamefont {Wang}\ \emph
  {et~al.}(2023{\natexlab{b}})\citenamefont {Wang}, \citenamefont {Gao},
  \citenamefont {Zhu}, \citenamefont {Ren}, \citenamefont {Hu}, \citenamefont
  {Sun}, \citenamefont {Ding}, \citenamefont {Xia},\ and\ \citenamefont
  {Li}}]{Wang2023}%
  \BibitemOpen
  \bibfield  {author} {\bibinfo {author} {\bibfnamefont {X.}~\bibnamefont
  {Wang}}, \bibinfo {author} {\bibfnamefont {Z.}~\bibnamefont {Gao}}, \bibinfo
  {author} {\bibfnamefont {G.}~\bibnamefont {Zhu}}, \bibinfo {author}
  {\bibfnamefont {J.}~\bibnamefont {Ren}}, \bibinfo {author} {\bibfnamefont
  {L.}~\bibnamefont {Hu}}, \bibinfo {author} {\bibfnamefont {J.}~\bibnamefont
  {Sun}}, \bibinfo {author} {\bibfnamefont {X.}~\bibnamefont {Ding}}, \bibinfo
  {author} {\bibfnamefont {Y.}~\bibnamefont {Xia}},\ and\ \bibinfo {author}
  {\bibfnamefont {B.}~\bibnamefont {Li}},\ }\bibfield  {title} {\bibinfo
  {title} {Role of high-order anharmonicity and off-diagonal terms in thermal
  conductivity: A case study of multiphase {CsPbBr$_3$}},\ }\href
  {https://doi.org/10.1103/physrevb.107.214308} {\bibfield  {journal} {\bibinfo
   {journal} {Phys. Rev. B}\ }\textbf {\bibinfo {volume} {107}},\ \bibinfo
  {pages} {214308} (\bibinfo {year} {2023}{\natexlab{b}})}\BibitemShut
  {NoStop}%
\bibitem [{\citenamefont {Liao}\ \emph {et~al.}(2015)\citenamefont {Liao},
  \citenamefont {Qiu}, \citenamefont {Zhou}, \citenamefont {Huberman},
  \citenamefont {Esfarjani},\ and\ \citenamefont {Chen}}]{Liao2015}%
  \BibitemOpen
  \bibfield  {author} {\bibinfo {author} {\bibfnamefont {B.}~\bibnamefont
  {Liao}}, \bibinfo {author} {\bibfnamefont {B.}~\bibnamefont {Qiu}}, \bibinfo
  {author} {\bibfnamefont {J.}~\bibnamefont {Zhou}}, \bibinfo {author}
  {\bibfnamefont {S.}~\bibnamefont {Huberman}}, \bibinfo {author}
  {\bibfnamefont {K.}~\bibnamefont {Esfarjani}},\ and\ \bibinfo {author}
  {\bibfnamefont {G.}~\bibnamefont {Chen}},\ }\bibfield  {title} {\bibinfo
  {title} {Significant reduction of lattice thermal conductivity by the
  electron-phonon interaction in silicon with high carrier concentrations: A
  first-principles study},\ }\href
  {https://doi.org/10.1103/physrevlett.114.115901} {\bibfield  {journal}
  {\bibinfo  {journal} {Phys. Rev. Lett.}\ }\textbf {\bibinfo {volume} {114}},\
  \bibinfo {pages} {115901} (\bibinfo {year} {2015})}\BibitemShut {NoStop}%
\bibitem [{\citenamefont {Jain}\ and\ \citenamefont
  {McGaughey}(2016)}]{Jain2016}%
  \BibitemOpen
  \bibfield  {author} {\bibinfo {author} {\bibfnamefont {A.}~\bibnamefont
  {Jain}}\ and\ \bibinfo {author} {\bibfnamefont {A.~J.~H.}\ \bibnamefont
  {McGaughey}},\ }\bibfield  {title} {\bibinfo {title} {Thermal transport by
  phonons and electrons in aluminum, silver, and gold from first principles},\
  }\href {https://doi.org/10.1103/physrevb.93.081206} {\bibfield  {journal}
  {\bibinfo  {journal} {Phys. Rev. B}\ }\textbf {\bibinfo {volume} {93}},\
  \bibinfo {pages} {081206} (\bibinfo {year} {2016})}\BibitemShut {NoStop}%
\bibitem [{\citenamefont {Wang}\ \emph {et~al.}(2016)\citenamefont {Wang},
  \citenamefont {Lu},\ and\ \citenamefont {Ruan}}]{Wang2016}%
  \BibitemOpen
  \bibfield  {author} {\bibinfo {author} {\bibfnamefont {Y.}~\bibnamefont
  {Wang}}, \bibinfo {author} {\bibfnamefont {Z.}~\bibnamefont {Lu}},\ and\
  \bibinfo {author} {\bibfnamefont {X.}~\bibnamefont {Ruan}},\ }\bibfield
  {title} {\bibinfo {title} {First principles calculation of lattice thermal
  conductivity of metals considering phonon-phonon and phonon-electron
  scattering},\ }\href {https://doi.org/10.1063/1.4953366} {\bibfield
  {journal} {\bibinfo  {journal} {J. Appl. Phys.}\ }\textbf {\bibinfo {volume}
  {119}},\ \bibinfo {pages} {225109} (\bibinfo {year} {2016})}\BibitemShut
  {NoStop}%
\bibitem [{\citenamefont {Yang}\ \emph {et~al.}(2002)\citenamefont {Yang},
  \citenamefont {Morelli}, \citenamefont {Meisner}, \citenamefont {Chen},
  \citenamefont {Dyck},\ and\ \citenamefont {Uher}}]{Yang2002}%
  \BibitemOpen
  \bibfield  {author} {\bibinfo {author} {\bibfnamefont {J.}~\bibnamefont
  {Yang}}, \bibinfo {author} {\bibfnamefont {D.~T.}\ \bibnamefont {Morelli}},
  \bibinfo {author} {\bibfnamefont {G.~P.}\ \bibnamefont {Meisner}}, \bibinfo
  {author} {\bibfnamefont {W.}~\bibnamefont {Chen}}, \bibinfo {author}
  {\bibfnamefont {J.~S.}\ \bibnamefont {Dyck}},\ and\ \bibinfo {author}
  {\bibfnamefont {C.}~\bibnamefont {Uher}},\ }\bibfield  {title} {\bibinfo
  {title} {Influence of electron-phonon interaction on the lattice thermal
  conductivity of {Co$_{1-x}$Ni$_x$Sb$_3$}},\ }\href
  {https://doi.org/10.1103/physrevb.65.094115} {\bibfield  {journal} {\bibinfo
  {journal} {Phys. Rev. B}\ }\textbf {\bibinfo {volume} {65}},\ \bibinfo
  {pages} {094115} (\bibinfo {year} {2002})}\BibitemShut {NoStop}%
\bibitem [{\citenamefont {Kundu}\ \emph {et~al.}(2021)\citenamefont {Kundu},
  \citenamefont {Yang}, \citenamefont {Ma}, \citenamefont {Feng}, \citenamefont
  {Carrete}, \citenamefont {Ruan}, \citenamefont {Madsen},\ and\ \citenamefont
  {Li}}]{Kundu2021}%
  \BibitemOpen
  \bibfield  {author} {\bibinfo {author} {\bibfnamefont {A.}~\bibnamefont
  {Kundu}}, \bibinfo {author} {\bibfnamefont {X.}~\bibnamefont {Yang}},
  \bibinfo {author} {\bibfnamefont {J.}~\bibnamefont {Ma}}, \bibinfo {author}
  {\bibfnamefont {T.}~\bibnamefont {Feng}}, \bibinfo {author} {\bibfnamefont
  {J.}~\bibnamefont {Carrete}}, \bibinfo {author} {\bibfnamefont
  {X.}~\bibnamefont {Ruan}}, \bibinfo {author} {\bibfnamefont {G.~K.}\
  \bibnamefont {Madsen}},\ and\ \bibinfo {author} {\bibfnamefont
  {W.}~\bibnamefont {Li}},\ }\bibfield  {title} {\bibinfo {title} {Ultrahigh
  thermal conductivity of $\theta$-phase tantalum nitride},\ }\href
  {https://doi.org/10.1103/physrevlett.126.115901} {\bibfield  {journal}
  {\bibinfo  {journal} {Phys. Rev. Lett.}\ }\textbf {\bibinfo {volume} {126}},\
  \bibinfo {pages} {115901} (\bibinfo {year} {2021})}\BibitemShut {NoStop}%
\bibitem [{\citenamefont {Wijeyesekera}\ and\ \citenamefont
  {Hoffmann}(1984)}]{Wijeyesekera1984}%
  \BibitemOpen
  \bibfield  {author} {\bibinfo {author} {\bibfnamefont {S.~D.}\ \bibnamefont
  {Wijeyesekera}}\ and\ \bibinfo {author} {\bibfnamefont {R.}~\bibnamefont
  {Hoffmann}},\ }\bibfield  {title} {\bibinfo {title} {Transition metal
  carbides. a comparison of bonding in extended and molecular interstitial
  carbides},\ }\href {https://doi.org/10.1021/om00085a001} {\bibfield
  {journal} {\bibinfo  {journal} {Organometallics}\ }\textbf {\bibinfo {volume}
  {3}},\ \bibinfo {pages} {949} (\bibinfo {year} {1984})}\BibitemShut {NoStop}%
\bibitem [{\citenamefont {Zhao}\ \emph {et~al.}(2014)\citenamefont {Zhao},
  \citenamefont {Sui}, \citenamefont {Tang}, \citenamefont {Lan}, \citenamefont
  {Jie}, \citenamefont {Kraemer}, \citenamefont {McEnaney}, \citenamefont
  {Guloy}, \citenamefont {Chen},\ and\ \citenamefont {Ren}}]{Zhao2014a}%
  \BibitemOpen
  \bibfield  {author} {\bibinfo {author} {\bibfnamefont {H.}~\bibnamefont
  {Zhao}}, \bibinfo {author} {\bibfnamefont {J.}~\bibnamefont {Sui}}, \bibinfo
  {author} {\bibfnamefont {Z.}~\bibnamefont {Tang}}, \bibinfo {author}
  {\bibfnamefont {Y.}~\bibnamefont {Lan}}, \bibinfo {author} {\bibfnamefont
  {Q.}~\bibnamefont {Jie}}, \bibinfo {author} {\bibfnamefont {D.}~\bibnamefont
  {Kraemer}}, \bibinfo {author} {\bibfnamefont {K.}~\bibnamefont {McEnaney}},
  \bibinfo {author} {\bibfnamefont {A.}~\bibnamefont {Guloy}}, \bibinfo
  {author} {\bibfnamefont {G.}~\bibnamefont {Chen}},\ and\ \bibinfo {author}
  {\bibfnamefont {Z.}~\bibnamefont {Ren}},\ }\bibfield  {title} {\bibinfo
  {title} {High thermoelectric performance of {MgAgSb}-based materials},\
  }\href {https://doi.org/10.1016/j.nanoen.2014.04.012} {\bibfield  {journal}
  {\bibinfo  {journal} {Nano Energy}\ }\textbf {\bibinfo {volume} {7}},\
  \bibinfo {pages} {97} (\bibinfo {year} {2014})}\BibitemShut {NoStop}%
\bibitem [{\citenamefont {Kirkham}\ \emph {et~al.}(2012)\citenamefont
  {Kirkham}, \citenamefont {dos Santos}, \citenamefont {Rawn}, \citenamefont
  {Lara-Curzio}, \citenamefont {Sharp},\ and\ \citenamefont
  {Thompson}}]{Kirkham2012}%
  \BibitemOpen
  \bibfield  {author} {\bibinfo {author} {\bibfnamefont {M.~J.}\ \bibnamefont
  {Kirkham}}, \bibinfo {author} {\bibfnamefont {A.~M.}\ \bibnamefont {dos
  Santos}}, \bibinfo {author} {\bibfnamefont {C.~J.}\ \bibnamefont {Rawn}},
  \bibinfo {author} {\bibfnamefont {E.}~\bibnamefont {Lara-Curzio}}, \bibinfo
  {author} {\bibfnamefont {J.~W.}\ \bibnamefont {Sharp}},\ and\ \bibinfo
  {author} {\bibfnamefont {A.~J.}\ \bibnamefont {Thompson}},\ }\bibfield
  {title} {\bibinfo {title} {$abinitio$ determination of crystal structures of
  the thermoelectric material {MgAgSb}},\ }\href
  {https://doi.org/10.1103/physrevb.85.144120} {\bibfield  {journal} {\bibinfo
  {journal} {Phys. Rev. B}\ }\textbf {\bibinfo {volume} {85}},\ \bibinfo
  {pages} {144120} (\bibinfo {year} {2012})}\BibitemShut {NoStop}%
\bibitem [{\citenamefont {Li}\ \emph {et~al.}(2025)\citenamefont {Li},
  \citenamefont {Wang}, \citenamefont {Wu}, \citenamefont {Li}, \citenamefont
  {Wang}, \citenamefont {Wu}, \citenamefont {Hu},\ and\ \citenamefont
  {Mori}}]{Li2025}%
  \BibitemOpen
  \bibfield  {author} {\bibinfo {author} {\bibfnamefont {A.}~\bibnamefont
  {Li}}, \bibinfo {author} {\bibfnamefont {L.}~\bibnamefont {Wang}}, \bibinfo
  {author} {\bibfnamefont {X.}~\bibnamefont {Wu}}, \bibinfo {author}
  {\bibfnamefont {J.}~\bibnamefont {Li}}, \bibinfo {author} {\bibfnamefont
  {X.}~\bibnamefont {Wang}}, \bibinfo {author} {\bibfnamefont {G.}~\bibnamefont
  {Wu}}, \bibinfo {author} {\bibfnamefont {Z.}~\bibnamefont {Hu}},\ and\
  \bibinfo {author} {\bibfnamefont {T.}~\bibnamefont {Mori}},\ }\bibfield
  {title} {\bibinfo {title} {Semiconductor-metal transition powers
  high-efficiency {MgAgSb} thermoelectrics},\ }\href
  {https://doi.org/10.1126/sciadv.adx7115} {\bibfield  {journal} {\bibinfo
  {journal} {Sci. Adv.}\ }\textbf {\bibinfo {volume} {11}},\ \bibinfo {pages}
  {eadx7115} (\bibinfo {year} {2025})}\BibitemShut {NoStop}%
\bibitem [{\citenamefont {Xie}\ \emph {et~al.}(2023)\citenamefont {Xie},
  \citenamefont {Yin}, \citenamefont {Yu}, \citenamefont {Peng}, \citenamefont
  {Song}, \citenamefont {Ying}, \citenamefont {Cai}, \citenamefont {Sun},
  \citenamefont {Shi}, \citenamefont {Wu}, \citenamefont {Qu}, \citenamefont
  {Guo}, \citenamefont {Cai}, \citenamefont {Wu}, \citenamefont {Zhang},
  \citenamefont {Nielsch}, \citenamefont {Ren}, \citenamefont {Liu},\ and\
  \citenamefont {Sui}}]{Xie2023a}%
  \BibitemOpen
  \bibfield  {author} {\bibinfo {author} {\bibfnamefont {L.}~\bibnamefont
  {Xie}}, \bibinfo {author} {\bibfnamefont {L.}~\bibnamefont {Yin}}, \bibinfo
  {author} {\bibfnamefont {Y.}~\bibnamefont {Yu}}, \bibinfo {author}
  {\bibfnamefont {G.}~\bibnamefont {Peng}}, \bibinfo {author} {\bibfnamefont
  {S.}~\bibnamefont {Song}}, \bibinfo {author} {\bibfnamefont {P.}~\bibnamefont
  {Ying}}, \bibinfo {author} {\bibfnamefont {S.}~\bibnamefont {Cai}}, \bibinfo
  {author} {\bibfnamefont {Y.}~\bibnamefont {Sun}}, \bibinfo {author}
  {\bibfnamefont {W.}~\bibnamefont {Shi}}, \bibinfo {author} {\bibfnamefont
  {H.}~\bibnamefont {Wu}}, \bibinfo {author} {\bibfnamefont {N.}~\bibnamefont
  {Qu}}, \bibinfo {author} {\bibfnamefont {F.}~\bibnamefont {Guo}}, \bibinfo
  {author} {\bibfnamefont {W.}~\bibnamefont {Cai}}, \bibinfo {author}
  {\bibfnamefont {H.}~\bibnamefont {Wu}}, \bibinfo {author} {\bibfnamefont
  {Q.}~\bibnamefont {Zhang}}, \bibinfo {author} {\bibfnamefont
  {K.}~\bibnamefont {Nielsch}}, \bibinfo {author} {\bibfnamefont
  {Z.}~\bibnamefont {Ren}}, \bibinfo {author} {\bibfnamefont {Z.}~\bibnamefont
  {Liu}},\ and\ \bibinfo {author} {\bibfnamefont {J.}~\bibnamefont {Sui}},\
  }\bibfield  {title} {\bibinfo {title} {Screening strategy for developing
  thermoelectric interface materials},\ }\href
  {https://doi.org/10.1126/science.adg8392} {\bibfield  {journal} {\bibinfo
  {journal} {Science}\ }\textbf {\bibinfo {volume} {382}},\ \bibinfo {pages}
  {921} (\bibinfo {year} {2023})}\BibitemShut {NoStop}%
\bibitem [{\citenamefont {Liu}\ \emph {et~al.}(2018)\citenamefont {Liu},
  \citenamefont {Mao}, \citenamefont {Sui},\ and\ \citenamefont
  {Ren}}]{Liu2018a}%
  \BibitemOpen
  \bibfield  {author} {\bibinfo {author} {\bibfnamefont {Z.}~\bibnamefont
  {Liu}}, \bibinfo {author} {\bibfnamefont {J.}~\bibnamefont {Mao}}, \bibinfo
  {author} {\bibfnamefont {J.}~\bibnamefont {Sui}},\ and\ \bibinfo {author}
  {\bibfnamefont {Z.}~\bibnamefont {Ren}},\ }\bibfield  {title} {\bibinfo
  {title} {High thermoelectric performance of $\alpha$-{MgAgSb} for power
  generation},\ }\href {https://doi.org/10.1039/c7ee02504a} {\bibfield
  {journal} {\bibinfo  {journal} {Energy Environ. Sci.}\ }\textbf {\bibinfo
  {volume} {11}},\ \bibinfo {pages} {23} (\bibinfo {year} {2018})}\BibitemShut
  {NoStop}%
\bibitem [{\citenamefont {Huang}\ \emph {et~al.}(2023)\citenamefont {Huang},
  \citenamefont {Lei}, \citenamefont {Chen}, \citenamefont {Zhou},
  \citenamefont {Dong}, \citenamefont {Yang}, \citenamefont {Gao},
  \citenamefont {Wei}, \citenamefont {Zhao},\ and\ \citenamefont
  {Shi}}]{Huang2023}%
  \BibitemOpen
  \bibfield  {author} {\bibinfo {author} {\bibfnamefont {Y.}~\bibnamefont
  {Huang}}, \bibinfo {author} {\bibfnamefont {J.}~\bibnamefont {Lei}}, \bibinfo
  {author} {\bibfnamefont {H.}~\bibnamefont {Chen}}, \bibinfo {author}
  {\bibfnamefont {Z.}~\bibnamefont {Zhou}}, \bibinfo {author} {\bibfnamefont
  {H.}~\bibnamefont {Dong}}, \bibinfo {author} {\bibfnamefont {S.}~\bibnamefont
  {Yang}}, \bibinfo {author} {\bibfnamefont {H.}~\bibnamefont {Gao}}, \bibinfo
  {author} {\bibfnamefont {T.-R.}\ \bibnamefont {Wei}}, \bibinfo {author}
  {\bibfnamefont {K.}~\bibnamefont {Zhao}},\ and\ \bibinfo {author}
  {\bibfnamefont {X.}~\bibnamefont {Shi}},\ }\bibfield  {title} {\bibinfo
  {title} {Intrinsically high thermoelectric performance in
  near-room-temperature $\alpha$-{MgAgSb} materials},\ }\href
  {https://doi.org/10.1016/j.actamat.2023.118847} {\bibfield  {journal}
  {\bibinfo  {journal} {Acta Mater.}\ }\textbf {\bibinfo {volume} {249}},\
  \bibinfo {pages} {118847} (\bibinfo {year} {2023})}\BibitemShut {NoStop}%
\bibitem [{\citenamefont {Back}\ \emph {et~al.}(2025)\citenamefont {Back},
  \citenamefont {Meikle},\ and\ \citenamefont {Mori}}]{Back2025}%
  \BibitemOpen
  \bibfield  {author} {\bibinfo {author} {\bibfnamefont {S.~Y.}\ \bibnamefont
  {Back}}, \bibinfo {author} {\bibfnamefont {S.}~\bibnamefont {Meikle}},\ and\
  \bibinfo {author} {\bibfnamefont {T.}~\bibnamefont {Mori}},\ }\bibfield
  {title} {\bibinfo {title} {Comprehensive study of $\alpha$-{MgAgSb}:
  Microstructure, carrier transport properties, and thermoelectric performance
  under ball milling techniques},\ }\href
  {https://doi.org/10.1016/j.jmst.2024.11.061} {\bibfield  {journal} {\bibinfo
  {journal} {J. Mater. Sci. Technol.}\ }\textbf {\bibinfo {volume} {227}},\
  \bibinfo {pages} {57} (\bibinfo {year} {2025})}\BibitemShut {NoStop}%
\bibitem [{\citenamefont {Ying}\ \emph {et~al.}(2015)\citenamefont {Ying},
  \citenamefont {Liu}, \citenamefont {Fu}, \citenamefont {Yue}, \citenamefont
  {Xie}, \citenamefont {Zhao}, \citenamefont {Zhang},\ and\ \citenamefont
  {Zhu}}]{Ying2015}%
  \BibitemOpen
  \bibfield  {author} {\bibinfo {author} {\bibfnamefont {P.}~\bibnamefont
  {Ying}}, \bibinfo {author} {\bibfnamefont {X.}~\bibnamefont {Liu}}, \bibinfo
  {author} {\bibfnamefont {C.}~\bibnamefont {Fu}}, \bibinfo {author}
  {\bibfnamefont {X.}~\bibnamefont {Yue}}, \bibinfo {author} {\bibfnamefont
  {H.}~\bibnamefont {Xie}}, \bibinfo {author} {\bibfnamefont {X.}~\bibnamefont
  {Zhao}}, \bibinfo {author} {\bibfnamefont {W.}~\bibnamefont {Zhang}},\ and\
  \bibinfo {author} {\bibfnamefont {T.}~\bibnamefont {Zhu}},\ }\bibfield
  {title} {\bibinfo {title} {High performance $\alpha$-{MgAgSb} thermoelectric
  materials for low temperature power generation},\ }\href
  {https://doi.org/10.1021/cm5041826} {\bibfield  {journal} {\bibinfo
  {journal} {Chem. Mater.}\ }\textbf {\bibinfo {volume} {27}},\ \bibinfo
  {pages} {909} (\bibinfo {year} {2015})}\BibitemShut {NoStop}%
\bibitem [{\citenamefont {Mi}\ \emph {et~al.}(2017)\citenamefont {Mi},
  \citenamefont {Ying}, \citenamefont {Sist}, \citenamefont {Reardon},
  \citenamefont {Zhang}, \citenamefont {Zhu}, \citenamefont {Zhao},\ and\
  \citenamefont {Iversen}}]{Mi2017}%
  \BibitemOpen
  \bibfield  {author} {\bibinfo {author} {\bibfnamefont {J.-L.}\ \bibnamefont
  {Mi}}, \bibinfo {author} {\bibfnamefont {P.-J.}\ \bibnamefont {Ying}},
  \bibinfo {author} {\bibfnamefont {M.}~\bibnamefont {Sist}}, \bibinfo {author}
  {\bibfnamefont {H.}~\bibnamefont {Reardon}}, \bibinfo {author} {\bibfnamefont
  {P.}~\bibnamefont {Zhang}}, \bibinfo {author} {\bibfnamefont {T.-J.}\
  \bibnamefont {Zhu}}, \bibinfo {author} {\bibfnamefont {X.-B.}\ \bibnamefont
  {Zhao}},\ and\ \bibinfo {author} {\bibfnamefont {B.~B.}\ \bibnamefont
  {Iversen}},\ }\bibfield  {title} {\bibinfo {title} {Elaborating the crystal
  structures of {MgAgSb} thermoelectric compound: Polymorphs and atomic
  disorders},\ }\href {https://doi.org/10.1021/acs.chemmater.7b01768}
  {\bibfield  {journal} {\bibinfo  {journal} {Chem. Mater.}\ }\textbf {\bibinfo
  {volume} {29}},\ \bibinfo {pages} {6378} (\bibinfo {year}
  {2017})}\BibitemShut {NoStop}%
\bibitem [{\citenamefont {Kresse}\ and\ \citenamefont
  {Furthmüller}(1996)}]{Kresse1996a}%
  \BibitemOpen
  \bibfield  {author} {\bibinfo {author} {\bibfnamefont {G.}~\bibnamefont
  {Kresse}}\ and\ \bibinfo {author} {\bibfnamefont {J.}~\bibnamefont
  {Furthmüller}},\ }\bibfield  {title} {\bibinfo {title} {Efficient iterative
  schemes forab initio total-energy calculations using a plane-wave basis
  set},\ }\href {https://doi.org/10.1103/physrevb.54.11169} {\bibfield
  {journal} {\bibinfo  {journal} {Phys. Rev. B}\ }\textbf {\bibinfo {volume}
  {54}},\ \bibinfo {pages} {11169} (\bibinfo {year} {1996})}\BibitemShut
  {NoStop}%
\bibitem [{\citenamefont {Perdew}\ \emph {et~al.}(2008)\citenamefont {Perdew},
  \citenamefont {Ruzsinszky}, \citenamefont {Csonka}, \citenamefont {Vydrov},
  \citenamefont {Scuseria}, \citenamefont {Constantin}, \citenamefont {Zhou},\
  and\ \citenamefont {Burke}}]{Perdew2008}%
  \BibitemOpen
  \bibfield  {author} {\bibinfo {author} {\bibfnamefont {J.~P.}\ \bibnamefont
  {Perdew}}, \bibinfo {author} {\bibfnamefont {A.}~\bibnamefont {Ruzsinszky}},
  \bibinfo {author} {\bibfnamefont {G.~I.}\ \bibnamefont {Csonka}}, \bibinfo
  {author} {\bibfnamefont {O.~A.}\ \bibnamefont {Vydrov}}, \bibinfo {author}
  {\bibfnamefont {G.~E.}\ \bibnamefont {Scuseria}}, \bibinfo {author}
  {\bibfnamefont {L.~A.}\ \bibnamefont {Constantin}}, \bibinfo {author}
  {\bibfnamefont {X.}~\bibnamefont {Zhou}},\ and\ \bibinfo {author}
  {\bibfnamefont {K.}~\bibnamefont {Burke}},\ }\bibfield  {title} {\bibinfo
  {title} {Restoring the density-gradient expansion for exchange in solids and
  surfaces},\ }\href {https://doi.org/10.1103/physrevlett.100.136406}
  {\bibfield  {journal} {\bibinfo  {journal} {Phys. Rev. Lett.}\ }\textbf
  {\bibinfo {volume} {100}},\ \bibinfo {pages} {136406} (\bibinfo {year}
  {2008})}\BibitemShut {NoStop}%
\bibitem [{\citenamefont {Hellman}\ and\ \citenamefont
  {Abrikosov}(2013)}]{Hellman2013}%
  \BibitemOpen
  \bibfield  {author} {\bibinfo {author} {\bibfnamefont {O.}~\bibnamefont
  {Hellman}}\ and\ \bibinfo {author} {\bibfnamefont {I.~A.}\ \bibnamefont
  {Abrikosov}},\ }\bibfield  {title} {\bibinfo {title} {Temperature-dependent
  effective third-order interatomic force constants from first principles},\
  }\href {https://doi.org/10.1103/physrevb.88.144301} {\bibfield  {journal}
  {\bibinfo  {journal} {Phys. Rev. B}\ }\textbf {\bibinfo {volume} {88}},\
  \bibinfo {pages} {144301} (\bibinfo {year} {2013})}\BibitemShut {NoStop}%
\bibitem [{\citenamefont {Knoop}\ \emph {et~al.}(2024)\citenamefont {Knoop},
  \citenamefont {Shulumba}, \citenamefont {Castellano}, \citenamefont
  {Batista}, \citenamefont {Farris}, \citenamefont {Verstraete}, \citenamefont
  {Heine}, \citenamefont {Broido}, \citenamefont {Kim}, \citenamefont
  {Klarbring}, \citenamefont {Abrikosov}, \citenamefont {Simak},\ and\
  \citenamefont {Hellman}}]{Knoop2024}%
  \BibitemOpen
  \bibfield  {author} {\bibinfo {author} {\bibfnamefont {F.}~\bibnamefont
  {Knoop}}, \bibinfo {author} {\bibfnamefont {N.}~\bibnamefont {Shulumba}},
  \bibinfo {author} {\bibfnamefont {A.}~\bibnamefont {Castellano}}, \bibinfo
  {author} {\bibfnamefont {J.~P.~A.}\ \bibnamefont {Batista}}, \bibinfo
  {author} {\bibfnamefont {R.}~\bibnamefont {Farris}}, \bibinfo {author}
  {\bibfnamefont {M.~J.}\ \bibnamefont {Verstraete}}, \bibinfo {author}
  {\bibfnamefont {M.}~\bibnamefont {Heine}}, \bibinfo {author} {\bibfnamefont
  {D.}~\bibnamefont {Broido}}, \bibinfo {author} {\bibfnamefont {D.~S.}\
  \bibnamefont {Kim}}, \bibinfo {author} {\bibfnamefont {J.}~\bibnamefont
  {Klarbring}}, \bibinfo {author} {\bibfnamefont {I.~A.}\ \bibnamefont
  {Abrikosov}}, \bibinfo {author} {\bibfnamefont {S.~I.}\ \bibnamefont
  {Simak}},\ and\ \bibinfo {author} {\bibfnamefont {O.}~\bibnamefont
  {Hellman}},\ }\bibfield  {title} {\bibinfo {title} {Tdep: Temperature
  dependent effective potentials},\ }\href
  {https://doi.org/10.21105/joss.06150} {\bibfield  {journal} {\bibinfo
  {journal} {J. Open Source Softw.}\ }\textbf {\bibinfo {volume} {9}},\
  \bibinfo {pages} {6150} (\bibinfo {year} {2024})}\BibitemShut {NoStop}%
\bibitem [{\citenamefont {Li}\ \emph {et~al.}(2014)\citenamefont {Li},
  \citenamefont {Carrete}, \citenamefont {A.~Katcho},\ and\ \citenamefont
  {Mingo}}]{Li2014}%
  \BibitemOpen
  \bibfield  {author} {\bibinfo {author} {\bibfnamefont {W.}~\bibnamefont
  {Li}}, \bibinfo {author} {\bibfnamefont {J.}~\bibnamefont {Carrete}},
  \bibinfo {author} {\bibfnamefont {N.}~\bibnamefont {A.~Katcho}},\ and\
  \bibinfo {author} {\bibfnamefont {N.}~\bibnamefont {Mingo}},\ }\bibfield
  {title} {\bibinfo {title} {Shengbte: A solver of the boltzmann transport
  equation for phonons},\ }\href {https://doi.org/10.1016/j.cpc.2014.02.015}
  {\bibfield  {journal} {\bibinfo  {journal} {Comput. Phys. Commu.}\ }\textbf
  {\bibinfo {volume} {185}},\ \bibinfo {pages} {1747} (\bibinfo {year}
  {2014})}\BibitemShut {NoStop}%
\bibitem [{\citenamefont {Han}\ \emph {et~al.}(2022)\citenamefont {Han},
  \citenamefont {Yang}, \citenamefont {Li}, \citenamefont {Feng},\ and\
  \citenamefont {Ruan}}]{Han2022}%
  \BibitemOpen
  \bibfield  {author} {\bibinfo {author} {\bibfnamefont {Z.}~\bibnamefont
  {Han}}, \bibinfo {author} {\bibfnamefont {X.}~\bibnamefont {Yang}}, \bibinfo
  {author} {\bibfnamefont {W.}~\bibnamefont {Li}}, \bibinfo {author}
  {\bibfnamefont {T.}~\bibnamefont {Feng}},\ and\ \bibinfo {author}
  {\bibfnamefont {X.}~\bibnamefont {Ruan}},\ }\bibfield  {title} {\bibinfo
  {title} {Fourphonon: An extension module to shengbte for computing
  four-phonon scattering rates and thermal conductivity},\ }\href
  {https://doi.org/10.1016/j.cpc.2021.108179} {\bibfield  {journal} {\bibinfo
  {journal} {Comput. Phys. Commun.}\ }\textbf {\bibinfo {volume} {270}},\
  \bibinfo {pages} {108179} (\bibinfo {year} {2022})}\BibitemShut {NoStop}%
\bibitem [{\citenamefont {Guo}\ \emph {et~al.}(2024)\citenamefont {Guo},
  \citenamefont {Han}, \citenamefont {Feng}, \citenamefont {Lin},\ and\
  \citenamefont {Ruan}}]{Guo2024}%
  \BibitemOpen
  \bibfield  {author} {\bibinfo {author} {\bibfnamefont {Z.}~\bibnamefont
  {Guo}}, \bibinfo {author} {\bibfnamefont {Z.}~\bibnamefont {Han}}, \bibinfo
  {author} {\bibfnamefont {D.}~\bibnamefont {Feng}}, \bibinfo {author}
  {\bibfnamefont {G.}~\bibnamefont {Lin}},\ and\ \bibinfo {author}
  {\bibfnamefont {X.}~\bibnamefont {Ruan}},\ }\bibfield  {title} {\bibinfo
  {title} {Sampling-accelerated prediction of phonon scattering rates for
  converged thermal conductivity and radiative properties},\ }\href
  {https://doi.org/10.1038/s41524-024-01215-8} {\bibfield  {journal} {\bibinfo
  {journal} {npj Comput. Mater.}\ }\textbf {\bibinfo {volume} {10}},\ \bibinfo
  {pages} {31} (\bibinfo {year} {2024})}\BibitemShut {NoStop}%
\bibitem [{\citenamefont {Ponc{\'{e}}}\ \emph {et~al.}(2016)\citenamefont
  {Ponc{\'{e}}}, \citenamefont {Margine}, \citenamefont {Verdi},\ and\
  \citenamefont {Giustino}}]{Ponce2016}%
  \BibitemOpen
  \bibfield  {author} {\bibinfo {author} {\bibfnamefont {S.}~\bibnamefont
  {Ponc{\'{e}}}}, \bibinfo {author} {\bibfnamefont {E.}~\bibnamefont
  {Margine}}, \bibinfo {author} {\bibfnamefont {C.}~\bibnamefont {Verdi}},\
  and\ \bibinfo {author} {\bibfnamefont {F.}~\bibnamefont {Giustino}},\
  }\bibfield  {title} {\bibinfo {title} {{EPW}: Electron{\textendash}phonon
  coupling, transport and superconducting properties using maximally localized
  wannier functions},\ }\href {https://doi.org/10.1016/j.cpc.2016.07.028}
  {\bibfield  {journal} {\bibinfo  {journal} {Comput. Phys. Commun.}\ }\textbf
  {\bibinfo {volume} {209}},\ \bibinfo {pages} {116} (\bibinfo {year}
  {2016})}\BibitemShut {NoStop}%
\bibitem [{\citenamefont {Giannozzi}\ \emph {et~al.}(2009)\citenamefont
  {Giannozzi}, \citenamefont {Baroni}, \citenamefont {Bonini}, \citenamefont
  {Calandra}, \citenamefont {Car}, \citenamefont {Cavazzoni}, \citenamefont
  {Ceresoli}, \citenamefont {Chiarotti}, \citenamefont {Cococcioni},
  \citenamefont {Dabo}, \citenamefont {Corso}, \citenamefont {de~Gironcoli},
  \citenamefont {Fabris}, \citenamefont {Fratesi}, \citenamefont {Gebauer},
  \citenamefont {Gerstmann}, \citenamefont {Gougoussis}, \citenamefont
  {Kokalj}, \citenamefont {Lazzeri}, \citenamefont {Martin-Samos},
  \citenamefont {Marzari}, \citenamefont {Mauri}, \citenamefont {Mazzarello},
  \citenamefont {Paolini}, \citenamefont {Pasquarello}, \citenamefont
  {Paulatto}, \citenamefont {Sbraccia}, \citenamefont {Scandolo}, \citenamefont
  {Sclauzero}, \citenamefont {Seitsonen}, \citenamefont {Smogunov},
  \citenamefont {Umari},\ and\ \citenamefont {Wentzcovitch}}]{Giannozzi2009}%
  \BibitemOpen
  \bibfield  {author} {\bibinfo {author} {\bibfnamefont {P.}~\bibnamefont
  {Giannozzi}}, \bibinfo {author} {\bibfnamefont {S.}~\bibnamefont {Baroni}},
  \bibinfo {author} {\bibfnamefont {N.}~\bibnamefont {Bonini}}, \bibinfo
  {author} {\bibfnamefont {M.}~\bibnamefont {Calandra}}, \bibinfo {author}
  {\bibfnamefont {R.}~\bibnamefont {Car}}, \bibinfo {author} {\bibfnamefont
  {C.}~\bibnamefont {Cavazzoni}}, \bibinfo {author} {\bibfnamefont
  {D.}~\bibnamefont {Ceresoli}}, \bibinfo {author} {\bibfnamefont {G.~L.}\
  \bibnamefont {Chiarotti}}, \bibinfo {author} {\bibfnamefont {M.}~\bibnamefont
  {Cococcioni}}, \bibinfo {author} {\bibfnamefont {I.}~\bibnamefont {Dabo}},
  \bibinfo {author} {\bibfnamefont {A.~D.}\ \bibnamefont {Corso}}, \bibinfo
  {author} {\bibfnamefont {S.}~\bibnamefont {de~Gironcoli}}, \bibinfo {author}
  {\bibfnamefont {S.}~\bibnamefont {Fabris}}, \bibinfo {author} {\bibfnamefont
  {G.}~\bibnamefont {Fratesi}}, \bibinfo {author} {\bibfnamefont
  {R.}~\bibnamefont {Gebauer}}, \bibinfo {author} {\bibfnamefont
  {U.}~\bibnamefont {Gerstmann}}, \bibinfo {author} {\bibfnamefont
  {C.}~\bibnamefont {Gougoussis}}, \bibinfo {author} {\bibfnamefont
  {A.}~\bibnamefont {Kokalj}}, \bibinfo {author} {\bibfnamefont
  {M.}~\bibnamefont {Lazzeri}}, \bibinfo {author} {\bibfnamefont
  {L.}~\bibnamefont {Martin-Samos}}, \bibinfo {author} {\bibfnamefont
  {N.}~\bibnamefont {Marzari}}, \bibinfo {author} {\bibfnamefont
  {F.}~\bibnamefont {Mauri}}, \bibinfo {author} {\bibfnamefont
  {R.}~\bibnamefont {Mazzarello}}, \bibinfo {author} {\bibfnamefont
  {S.}~\bibnamefont {Paolini}}, \bibinfo {author} {\bibfnamefont
  {A.}~\bibnamefont {Pasquarello}}, \bibinfo {author} {\bibfnamefont
  {L.}~\bibnamefont {Paulatto}}, \bibinfo {author} {\bibfnamefont
  {C.}~\bibnamefont {Sbraccia}}, \bibinfo {author} {\bibfnamefont
  {S.}~\bibnamefont {Scandolo}}, \bibinfo {author} {\bibfnamefont
  {G.}~\bibnamefont {Sclauzero}}, \bibinfo {author} {\bibfnamefont {A.~P.}\
  \bibnamefont {Seitsonen}}, \bibinfo {author} {\bibfnamefont {A.}~\bibnamefont
  {Smogunov}}, \bibinfo {author} {\bibfnamefont {P.}~\bibnamefont {Umari}},\
  and\ \bibinfo {author} {\bibfnamefont {R.~M.}\ \bibnamefont {Wentzcovitch}},\
  }\bibfield  {title} {\bibinfo {title} {{QUANTUM} {ESPRESSO}: a modular and
  open-source software project for quantum simulations of materials},\ }\href
  {https://doi.org/10.1088/0953-8984/21/39/395502} {\bibfield  {journal}
  {\bibinfo  {journal} {J. Phys.:Condens. Matter}\ }\textbf {\bibinfo {volume}
  {21}},\ \bibinfo {pages} {395502} (\bibinfo {year} {2009})}\BibitemShut
  {NoStop}%
\bibitem [{\citenamefont {Hamann}(2013)}]{Hamann2013}%
  \BibitemOpen
  \bibfield  {author} {\bibinfo {author} {\bibfnamefont {D.~R.}\ \bibnamefont
  {Hamann}},\ }\bibfield  {title} {\bibinfo {title} {Optimized norm-conserving
  vanderbilt pseudopotentials},\ }\href
  {https://doi.org/10.1103/physrevb.88.085117} {\bibfield  {journal} {\bibinfo
  {journal} {Phys. Rev. B}\ }\textbf {\bibinfo {volume} {88}},\ \bibinfo
  {pages} {085117} (\bibinfo {year} {2013})}\BibitemShut {NoStop}%
\bibitem [{\citenamefont {Li}\ \emph {et~al.}(2012{\natexlab{a}})\citenamefont
  {Li}, \citenamefont {Mingo}, \citenamefont {Lindsay}, \citenamefont {Broido},
  \citenamefont {Stewart},\ and\ \citenamefont {Katcho}}]{Li2012b}%
  \BibitemOpen
  \bibfield  {author} {\bibinfo {author} {\bibfnamefont {W.}~\bibnamefont
  {Li}}, \bibinfo {author} {\bibfnamefont {N.}~\bibnamefont {Mingo}}, \bibinfo
  {author} {\bibfnamefont {L.}~\bibnamefont {Lindsay}}, \bibinfo {author}
  {\bibfnamefont {D.~A.}\ \bibnamefont {Broido}}, \bibinfo {author}
  {\bibfnamefont {D.~A.}\ \bibnamefont {Stewart}},\ and\ \bibinfo {author}
  {\bibfnamefont {N.~A.}\ \bibnamefont {Katcho}},\ }\bibfield  {title}
  {\bibinfo {title} {Thermal conductivity of diamond nanowires from first
  principles},\ }\href {https://doi.org/10.1103/physrevb.85.195436} {\bibfield
  {journal} {\bibinfo  {journal} {Phys. Rev. B}\ }\textbf {\bibinfo {volume}
  {85}},\ \bibinfo {pages} {195436} (\bibinfo {year}
  {2012}{\natexlab{a}})}\BibitemShut {NoStop}%
\bibitem [{\citenamefont {Li}\ \emph {et~al.}(2012{\natexlab{b}})\citenamefont
  {Li}, \citenamefont {Lindsay}, \citenamefont {Broido}, \citenamefont
  {Stewart},\ and\ \citenamefont {Mingo}}]{Li2012c}%
  \BibitemOpen
  \bibfield  {author} {\bibinfo {author} {\bibfnamefont {W.}~\bibnamefont
  {Li}}, \bibinfo {author} {\bibfnamefont {L.}~\bibnamefont {Lindsay}},
  \bibinfo {author} {\bibfnamefont {D.~A.}\ \bibnamefont {Broido}}, \bibinfo
  {author} {\bibfnamefont {D.~A.}\ \bibnamefont {Stewart}},\ and\ \bibinfo
  {author} {\bibfnamefont {N.}~\bibnamefont {Mingo}},\ }\bibfield  {title}
  {\bibinfo {title} {Thermal conductivity of bulk and nanowire
  {Mg$_2$Si$_x$Sn$_{1-x}$} alloys from first principles},\ }\href
  {https://doi.org/10.1103/physrevb.86.174307} {\bibfield  {journal} {\bibinfo
  {journal} {Phys. Rev. B}\ }\textbf {\bibinfo {volume} {86}},\ \bibinfo
  {pages} {174307} (\bibinfo {year} {2012}{\natexlab{b}})}\BibitemShut
  {NoStop}%
\bibitem [{\citenamefont {Ouyang}\ \emph {et~al.}(2025)\citenamefont {Ouyang},
  \citenamefont {Shen}, \citenamefont {Wang}, \citenamefont {Cheng},
  \citenamefont {Wang},\ and\ \citenamefont {Chen}}]{Ouyang2025}%
  \BibitemOpen
  \bibfield  {author} {\bibinfo {author} {\bibfnamefont {N.}~\bibnamefont
  {Ouyang}}, \bibinfo {author} {\bibfnamefont {D.}~\bibnamefont {Shen}},
  \bibinfo {author} {\bibfnamefont {C.}~\bibnamefont {Wang}}, \bibinfo {author}
  {\bibfnamefont {R.}~\bibnamefont {Cheng}}, \bibinfo {author} {\bibfnamefont
  {Q.}~\bibnamefont {Wang}},\ and\ \bibinfo {author} {\bibfnamefont
  {Y.}~\bibnamefont {Chen}},\ }\bibfield  {title} {\bibinfo {title} {Positive
  temperature-dependent thermal conductivity induced by wavelike phonons in
  complex {Ag}-based argyrodites},\ }\href
  {https://doi.org/10.1103/physrevb.111.064307} {\bibfield  {journal} {\bibinfo
   {journal} {Phys. Rev. B}\ }\textbf {\bibinfo {volume} {111}},\ \bibinfo
  {pages} {064307} (\bibinfo {year} {2025})}\BibitemShut {NoStop}%
\bibitem [{\citenamefont {Zhao}\ \emph {et~al.}(2025)\citenamefont {Zhao},
  \citenamefont {Zhang}, \citenamefont {Zhang}, \citenamefont {Shin},\ and\
  \citenamefont {Shen}}]{Zhao2025}%
  \BibitemOpen
  \bibfield  {author} {\bibinfo {author} {\bibfnamefont {Y.-M.}\ \bibnamefont
  {Zhao}}, \bibinfo {author} {\bibfnamefont {X.}~\bibnamefont {Zhang}},
  \bibinfo {author} {\bibfnamefont {C.}~\bibnamefont {Zhang}}, \bibinfo
  {author} {\bibfnamefont {S.}~\bibnamefont {Shin}},\ and\ \bibinfo {author}
  {\bibfnamefont {L.}~\bibnamefont {Shen}},\ }\bibfield  {title} {\bibinfo
  {title} {Dual-channel phonon transport in two-dimensional materials with low
  thermal conductivity},\ }\href {https://doi.org/10.1103/42b8-kpld} {\bibfield
   {journal} {\bibinfo  {journal} {Phys. Rev. B}\ }\textbf {\bibinfo {volume}
  {112}},\ \bibinfo {pages} {125406} (\bibinfo {year} {2025})}\BibitemShut
  {NoStop}%
\bibitem [{\citenamefont {Zheng}\ \emph {et~al.}(2024)\citenamefont {Zheng},
  \citenamefont {Lin}, \citenamefont {Lin}, \citenamefont {Hautier},
  \citenamefont {Guo},\ and\ \citenamefont {Huang}}]{Zheng2024}%
  \BibitemOpen
  \bibfield  {author} {\bibinfo {author} {\bibfnamefont {J.}~\bibnamefont
  {Zheng}}, \bibinfo {author} {\bibfnamefont {C.}~\bibnamefont {Lin}}, \bibinfo
  {author} {\bibfnamefont {C.}~\bibnamefont {Lin}}, \bibinfo {author}
  {\bibfnamefont {G.}~\bibnamefont {Hautier}}, \bibinfo {author} {\bibfnamefont
  {R.}~\bibnamefont {Guo}},\ and\ \bibinfo {author} {\bibfnamefont
  {B.}~\bibnamefont {Huang}},\ }\bibfield  {title} {\bibinfo {title}
  {Unravelling ultralow thermal conductivity in perovskite {Cs$_2$AgBiBr$_6$}:
  dominant wave-like phonon tunnelling and strong anharmonicity},\ }\href
  {https://doi.org/10.1038/s41524-024-01211-y} {\bibfield  {journal} {\bibinfo
  {journal} {npj Comput. Mater.}\ }\textbf {\bibinfo {volume} {10}},\ \bibinfo
  {pages} {30} (\bibinfo {year} {2024})}\BibitemShut {NoStop}%
\bibitem [{\citenamefont {Wu}\ \emph {et~al.}(2023)\citenamefont {Wu},
  \citenamefont {Chen}, \citenamefont {Fang}, \citenamefont {Ding},
  \citenamefont {Li}, \citenamefont {Xue}, \citenamefont {Shao}, \citenamefont
  {Zhang},\ and\ \citenamefont {Zhou}}]{Wu2023b}%
  \BibitemOpen
  \bibfield  {author} {\bibinfo {author} {\bibfnamefont {Y.}~\bibnamefont
  {Wu}}, \bibinfo {author} {\bibfnamefont {Y.}~\bibnamefont {Chen}}, \bibinfo
  {author} {\bibfnamefont {Z.}~\bibnamefont {Fang}}, \bibinfo {author}
  {\bibfnamefont {Y.}~\bibnamefont {Ding}}, \bibinfo {author} {\bibfnamefont
  {Q.}~\bibnamefont {Li}}, \bibinfo {author} {\bibfnamefont {K.}~\bibnamefont
  {Xue}}, \bibinfo {author} {\bibfnamefont {H.}~\bibnamefont {Shao}}, \bibinfo
  {author} {\bibfnamefont {H.}~\bibnamefont {Zhang}},\ and\ \bibinfo {author}
  {\bibfnamefont {L.}~\bibnamefont {Zhou}},\ }\bibfield  {title} {\bibinfo
  {title} {Ultralow lattice thermal transport and considerable wave-like phonon
  tunneling in chalcogenide perovskite {BaZrS$_3$}},\ }\href
  {https://doi.org/10.1021/acs.jpclett.3c02940} {\bibfield  {journal} {\bibinfo
   {journal} {J. Phys. Chem. Lett.}\ }\textbf {\bibinfo {volume} {14}},\
  \bibinfo {pages} {11465} (\bibinfo {year} {2023})}\BibitemShut {NoStop}%
\bibitem [{\citenamefont {Di~Lucente}\ \emph {et~al.}(2023)\citenamefont
  {Di~Lucente}, \citenamefont {Simoncelli},\ and\ \citenamefont
  {Marzari}}]{DiLucente2023}%
  \BibitemOpen
  \bibfield  {author} {\bibinfo {author} {\bibfnamefont {E.}~\bibnamefont
  {Di~Lucente}}, \bibinfo {author} {\bibfnamefont {M.}~\bibnamefont
  {Simoncelli}},\ and\ \bibinfo {author} {\bibfnamefont {N.}~\bibnamefont
  {Marzari}},\ }\bibfield  {title} {\bibinfo {title} {Crossover from boltzmann
  to wigner thermal transport in thermoelectric skutterudites},\ }\href
  {https://doi.org/10.1103/physrevresearch.5.033125} {\bibfield  {journal}
  {\bibinfo  {journal} {Phys. Rev. Res.}\ }\textbf {\bibinfo {volume} {5}},\
  \bibinfo {pages} {033125} (\bibinfo {year} {2023})}\BibitemShut {NoStop}%
\bibitem [{\citenamefont {Li}\ \emph {et~al.}(2024)\citenamefont {Li},
  \citenamefont {Chen}, \citenamefont {Lu}, \citenamefont {Fukui},
  \citenamefont {Yu}, \citenamefont {Li}, \citenamefont {Zhao}, \citenamefont
  {Wang}, \citenamefont {Wang},\ and\ \citenamefont {Hong}}]{Li2024}%
  \BibitemOpen
  \bibfield  {author} {\bibinfo {author} {\bibfnamefont {Y.}~\bibnamefont
  {Li}}, \bibinfo {author} {\bibfnamefont {J.}~\bibnamefont {Chen}}, \bibinfo
  {author} {\bibfnamefont {C.}~\bibnamefont {Lu}}, \bibinfo {author}
  {\bibfnamefont {H.}~\bibnamefont {Fukui}}, \bibinfo {author} {\bibfnamefont
  {X.}~\bibnamefont {Yu}}, \bibinfo {author} {\bibfnamefont {C.}~\bibnamefont
  {Li}}, \bibinfo {author} {\bibfnamefont {J.}~\bibnamefont {Zhao}}, \bibinfo
  {author} {\bibfnamefont {X.}~\bibnamefont {Wang}}, \bibinfo {author}
  {\bibfnamefont {W.}~\bibnamefont {Wang}},\ and\ \bibinfo {author}
  {\bibfnamefont {J.}~\bibnamefont {Hong}},\ }\bibfield  {title} {\bibinfo
  {title} {Multiphonon interaction and thermal conductivity in half-heusler
  {LuNiBi}},\ }\href {https://doi.org/10.1103/physrevb.109.174302} {\bibfield
  {journal} {\bibinfo  {journal} {Phys. Rev. B}\ }\textbf {\bibinfo {volume}
  {109}},\ \bibinfo {pages} {174302} (\bibinfo {year} {2024})}\BibitemShut
  {NoStop}%
\end{thebibliography}

\end{document}